\documentclass[12pt,letterpaper]{article}

\usepackage{graphicx}
\usepackage{booktabs}
\usepackage[dvipsnames]{xcolor}

\newcommand{\maxwidth}{\textwidth}

\usepackage{alltt}

\usepackage{setspace}
\usepackage{pdflscape}
\usepackage{array}
\usepackage{afterpage}
\usepackage{amsmath}
\usepackage{siunitx}
\usepackage{lineno}

\newcolumntype{M}[1]{>{\centering\arraybackslash}m{#1}}

\newcommand{\R}{{\mathcal R}}
\newcommand{\ie}{\emph{i.e., }}

\newcommand{\CFP}{{\texttt{\upshape CFP}}}

\usepackage{xparse}
\usepackage{placeins}
  \usepackage[letterpaper,margin=1in]{geometry}
  \usepackage{caption}
  \usepackage{natbib}
  \setcitestyle{numbers,square}
  \usepackage{hyperref}
  \usepackage[nameinlink,capitalize]{cleveref}
  
\usepackage{xurl}
\crefname{section}{\S}{\S\S}
\crefname{equation}{Equation}{Equations}
\crefname{figure}{Figure}{Figures}
\crefname{appendix}{Appendix}{Appendices}
\crefname{table}{Table}{Tables}

\newcommand{\mstitle}{Scarlet Fever Dynamics in 19th and 20th Century London}

\newcommand{\msauthorone}{Kevin H.\ Zhao}
\newcommand{\msauthortwo}{David J.\,D.\ Earn}
\newcommand{\mspdfauthors}{Kevin H. Zhao and David J. D. Earn}
\newcommand{\msaffiliationone}{Department of Mathematics \& Statistics}
\newcommand{\msaffiliationtwo}{M.\,G.\ DeGroote Institute for Infectious Disease Research}
\newcommand{\msuniversityaddress}{%
  McMaster University, 1280 Main Street West, Hamilton, Ontario, L8S 4K1, Canada%
}
\newcommand{\msfulladdress}{%
  \msaffiliationone{} and \msaffiliationtwo, \msuniversityaddress%
}

\newcommand{\msauthoroneorcid}{0009-0002-8658-1717}
\newcommand{\msauthoroneorcidurl}{https://orcid.org/0009-0002-8658-1717}
\newcommand{\msauthortwoorcid}{0000-0002-7562-1341}
\newcommand{\msauthortwoorcidurl}{https://orcid.org/0000-0002-7562-1341}

\newcommand{\msarticleauthors}{%
  \parbox{0.9\textwidth}{\centering
    \msauthorone{} and \msauthortwo\\[0.75ex]
    \small \msfulladdress\\[0.5ex]
    \small ORCID: \textbf{KHZ}, \href{\msauthoroneorcidurl}\msauthoroneorcid;
    \textbf{DJDE}, \href{\msauthortwoorcidurl}{\msauthortwoorcid}\\[0.5ex]
    \small Correspondence: \href{mailto:\mscorrespondingemail}{\mscorrespondingemail}%
  }%
}

\newcommand{\mscorrespondingemail}{earn@math.mcmaster.ca}

\newcommand{\mssubjectcategory}{Life Sciences--Mathematics interface}
\newcommand{\mssubjects}{biomathematics}
\newcommand{\mskeywords}{scarlet fever, London, SIR model, bifurcations, transition analysis}

\newcommand{\msfunding}{DJDE was supported by a Discovery Grant from the
  Natural Sciences and Engineering Research Council of Canada (NSERC).  KHZ
  was supported by NSERC and an Ontario Graduate Scholarship (OGS).}

\newcommand{\msacknowledgements}{We are grateful to all members of the
  Mac-Theobio research group for feedback on preliminary presentations, and
  especially to Steve Walker and Ben Bolker, the principal developers of
  \texttt{macpan2}, and Mikael Jagan, the developer of \texttt{fastbeta}.}

\newcommand{\msaiuse}{ChatGPT and OpenAI Codex assisted with the
  transcription of data from publicly available scans of historical
  tables for the period 1901--1939 and with language editing of the
  manuscript.  The authors take full responsibility for the final
  manuscript and data.}

  \title{\mstitle}
  \author{\msarticleauthors}
  \date{}
\hypersetup{%
  hidelinks,
  pdftitle={\mstitle},
  pdfauthor={\mspdfauthors},
  pdfsubject={\mssubjectcategory; \mssubjects},
  pdfkeywords={\mskeywords}%
}
\IfFileExists{upquote.sty}{\usepackage{upquote}}{}

\begin{document}

  \maketitle
  \begin{abstract}
    Weekly scarlet fever (SF) mortality records for London, UK, from 1842
to 1939, together with notified case records from 1901 to 1939, reveal
a strong annual epidemic pattern with peak prevalence in the autumn.
In addition to the annual epidemic pattern, this long time series
reveals a cyclical envelope with a period that lengthened over the
decades.  In particular, the period of the envelope increased from
about four years to about eight years between 1880 and 1920,
coinciding with a dramatic decline in pre-antibiotic-era SF deaths
(from about 2300/yr to about 80/yr).
We quantify the spectral features of the SF time series using a
wavelet transform, and attempt to explain why the frequency structure
changed over time, using a mechanistic mathematical model of SF
transmission dynamics.
We estimate the parameters of the model in part from the literature
and in part by fitting mortality and incidence jointly.
We use a mechanistic transition analysis---considering both
attractors and transients in model solutions---to relate the
observed changes in frequency structure to our estimated changes in
model parameters.
We find that almost all spectral evolution in the time series can be
explained by changes in the effective reproduction number and the
amplitude of seasonal forcing.  These changes reflect observed
variation in birth rates together with inferred changes in
transmission, which may in part have arisen through pathogen
evolution.

  \end{abstract}
  \par

  \newpage
  \section{Introduction}

Scarlet fever was one of the most feared childhood infections of the
19th century.  Between about 1820 and 1880, severe epidemics occurred
throughout Europe and North America, and in the middle of the century
scarlet fever caused approximately 10,000 deaths annually in England
and Wales~\citep{Hard93,SwedDont02}.  The burden fell mainly on young
children, among whom the disease could progress from its
characteristic sore throat and rash to death within a few
days~\citep{SwedDont02}.  In London, published weekly mortality and,
later, notified incidence counts
provide an unusually long and detailed record of these epidemics.
Such historical records provide an empirical backbone for mechanistic
mathematical modelling of transmission dynamics, making it possible to
study changes in epidemic patterns over decades or even
centuries~\citep{Earn09,KrylEarn20,Earn+25}.

Scarlet fever mortality changed dramatically in London during the 19th
and early 20th centuries.  Mortality was high and epidemics recurred
regularly through most of the 19th century, but the mortality rate
then declined by more than an order of magnitude, several decades
before effective antibacterial drugs became widely available in the
late 1930s and 1940s~\citep{Amin10}.  At the same time, the seasonal
and multi-year patterns of fluctuations in the mortality series
changed.  These changes raise a mechanistic question: can mathematical
modelling explain the changing epidemic patterns as dynamical
consequences of changes in epidemiological parameters?

Rapid qualitative changes in recurrent epidemic patterns are not
unique to scarlet fever.  Earn et al.~\citep{Earn+00,Earn09} showed
that slow changes in the recruitment of susceptible
individuals---through changes in birth rates or vaccination---could
move a seasonally forced measles model among attractors with different
epidemic periodicities, explaining observed transitions among annual,
multi-year and irregular epidemic patterns.  Bauch and
Earn~\citep{BaucEarn03a} subsequently applied this approach to several
childhood infections.  Attractors explained the dominant periodicities
of measles and chickenpox, but not those of rubella or whooping cough.
The missing periodicities arose from transient oscillations during
approach to an attractor; demographic stochasticity continually
excites these transients, allowing them to persist in observed
epidemic time series.  The same analysis also showed that what had
appeared to be irregularity in measles dynamics could reflect
stochastically sustained transients rather than stochastic switching
among coexisting attractors.  We use the term \emph{transition
analysis}~\citep{KrylEarn13} for the combined study of how slowly
changing epidemiological parameters alter attractors, their basins of
attraction and observable transient periodicities.  Applied to nearly
a century of New York City measles incidence, this approach explained
a sequence of observed changes in frequency structure with the same
seasonally forced model~\citep{HempEarn15}.

Several results support the use of deliberately simple models for
transition analysis.  For a fixed mean generation interval, suitably
parameterized SIR and SEIR models make nearly identical predictions of
transitions in measles dynamics, even when latent and infectious stage
durations have realistic distributions~\citep{KrylEarn13}.  Similarly,
representing seasonal variation in transmission due to school terms
explicitly can substantially improve a time-series fit without
appreciably changing the qualitative bifurcation
structure~\citep{HeEarn16}.  More generally, the key
bifurcations of the standard seasonally forced SIR model are invariant
to radically different seasonal forcing shapes when the forcing
amplitude is adjusted appropriately~\citep{PapsEarn19}.  Consequently,
for transition analysis, we can replace the seasonal forcing function
fitted for each epoch by a sinusoid whose amplitude is recalibrated by
matching the location of a reference period-doubling bifurcation.
This allows us to represent the inferred changes as a trajectory
through the two-parameter (basic reproduction number, seasonal forcing
amplitude) plane, following the approach used by Hempel and
Earn~\citep{HempEarn15}.  A separate question is whether the inferred
transmission trajectory depends strongly on the estimation method.
Fast reconstruction provides a computationally inexpensive alternative
to likelihood-based fitting and can therefore be used to determine
whether the inferred transmission pattern depends strongly on the
estimation method~\citep{Jaga+20}.

Scarlet fever provides a more stringent test of this framework than
measles.  The century-long, high-resolution record analysed here
consists mainly of weekly deaths; weekly notified scarlet fever case
reports did not appear in the RGWR until 1901.  Consequently, fitting the transmission model
requires an observation process that links modelled infection
incidence to deaths through the case-fatality proportion and the delay
from infection to death.  The pronounced mortality decline could
therefore reflect changes in infection incidence, fatality, reporting,
or some combination of these.  In addition, scarlet fever is caused by
the bacterium \emph{Streptococcus pyogenes}; repeat episodes occur,
and infection does not confer the simple, permanent immunity
characteristic of measles~\citep{Wess16,deDi17}.  Changes in the
virulence of circulating strains have also been proposed as an
explanation for the rise and decline of severe 19th-century
epidemics~\citep{SwedDont02}.  Weekly christening and birth records
extend back to 1664~\citep{KrylEarn20}, providing a direct historical
measure of changes in susceptible recruitment throughout the period
analysed.  The observed demographic series and the estimated changes
in mean and seasonal transmission define a trajectory through model
parameter space.  We ask whether the attractor and transient
periodicities predicted along this trajectory correspond to the
changing periodicities observed in the mortality record.

In this paper, we analyse weekly scarlet fever mortality in London
from 1842 to 1939 and notified cases from 1901 to 1939.  We first describe changes in mortality,
seasonality and frequency structure, using a continuous wavelet
transform to identify when dominant epidemic periodicities changed.
We then fit a seasonally forced SIR model with a mortality observation
process, allowing the case-fatality proportion, mean transmission rate
and seasonal forcing to vary over time.  We use a much faster
reconstruction method to test whether the inferred transmission
trajectory depends strongly on the estimation procedure.  We then
combine directly observed changes in birth rates with estimated
changes in mean and seasonal transmission to predict attractor and
transient periodicities, which we compare with the changing
periodicities identified by wavelet analysis.  We also assess whether
the main conclusions are robust to the principal uncertain components
of the analysis.

Our aim is not to reproduce every detail of the mortality and incidence records with
a minimal SIR model.  We begin with the simplest model that could
plausibly provide a mechanistic explanation; failures of that model
indicate where additional biological structure may be required.  We
find that transition analysis accounts for almost all of the prominent
periodicities from 1842 to 1939, but not the approximately five-year
fluctuations observed from 1842 to 1857, the four-year fluctuations
from 1925 to 1939, and potential two- and three-year cycles from 1842
to 1860 and 1890 to 1900, respectively.  The
agreement across most transitions extends transition analysis to a
disease and data type that violate several assumptions that are
relatively benign for measles and other childhood infections that have
been successfully explained by this method, while the unexplained
features identify where additional biological structure or historical
information may be required to explain scarlet fever dynamics.

\section{Scarlet fever biology}

Scarlet fever is caused by \emph{Streptococcus pyogenes}, one of the Group A Streptococci that primarily infect humans \citep{Wess16}. 
The link between streptococcal bacteria and scarlet fever was established in 1924 by George and Gladys Dick \citep[p.\,302]{DickDick24}, after which throat cultures began to be used as a diagnostic test for scarlet fever.

An early description of the clinical course of scarlet fever was given in 1825 by the American physician William Potts Dewees in \emph{A Treatise on the Physical and Medical Treatment of Children}:
\begin{quote}
  ``As the disease proceeds, the neck and lower jaw grow stiff, the tonsils swell, and become marked with specks, which degenerate into ulcers, covered with superficial, ash-coloured sloughs.
  The sloughs on the tonsils grow fouler; and the discharge from them, and the nostrils, becomes exceedingly acrid\dots
  When the disease is very malignant, death sometimes takes place on the third or fourth day;
  while in its milder forms, it [illness symptoms] will linger on to the second, or even third week'' \citep{Dewe1825}, as quoted by \citealp{RadiConn07}.
\end{quote}
Dewees' observations from 200 years ago are consistent with many of
the symptoms that are now considered characteristic scarlet fever: the
sore throat associated with \emph{S.~pyogenes} is caused by swelling
in the tonsils and posterior oropharynx, and the oropharynx can become
white due to the accumulation of blood during the immune response,
oozing yellow or white pus \citep[p.117]{Rose18}.  While the
characteristic red rash is almost always present, many other symptoms
of streptococcal pharyngitis also manifest \citep{Wess16}, including
fever, nausea, vomiting, and strawberry tongue \citep{RadiConn07}.
Although symptoms are common, infected individuals can vary widely in
presentation, with some showing no symptoms at all but acting as
infectious carriers \citep[p.\,159]{SwedDont02}.  For instance, a
longitudinal study conducted in Pittsburgh, Pennsylvania from 1998 to
2002 found that 13--18\% of children are asymptomatic carriers of
Group A Streptococci \citep[p.\,1214]{Mart+04}.

Estimates of parameters that characterize the natural history of
infection of SF are summarized in \cref{tab:Params}.  Ranges for the
incubation, latent and infectious periods are given in several
sources.  Since the latent period is typically shorter than the
incubation period, pre-symptomatic transmission is possible.  The
infectious period can be as long as three weeks.  However, treatment
with penicillin can reduce the infectious period to less than a day
\citep{Heym04}.

Newborns are generally found to be immune to scarlet fever during the
first year of life, likely due to maternal antibodies.  Immunity to
\emph{S.~pyogenes} is generally acquired by repeated infection (some
of which might be asymptomatic); in particular, developing scarlet
fever once does not guarantee immunity \citep[p.\,8]{DickDick34}.

\section{Data summary}

Weekly scarlet fever mortality data for London, England, are available
from 1830 to 1950, collected in the LBoM (1830--1845) and the RGWR
(1842--1950), together with weekly birth and all-cause mortality data.
Population estimates for London are also available from census data taken every 10 years (and estimated before the start of World War 2 in 1939) \citep{Demographia}.
We outline the characteristics of these datasets in the following subsections.

\subsection{SF in the London Bills of Mortality, 1696--1845}

\begin{figure}[p]
  \includegraphics[width=\maxwidth]{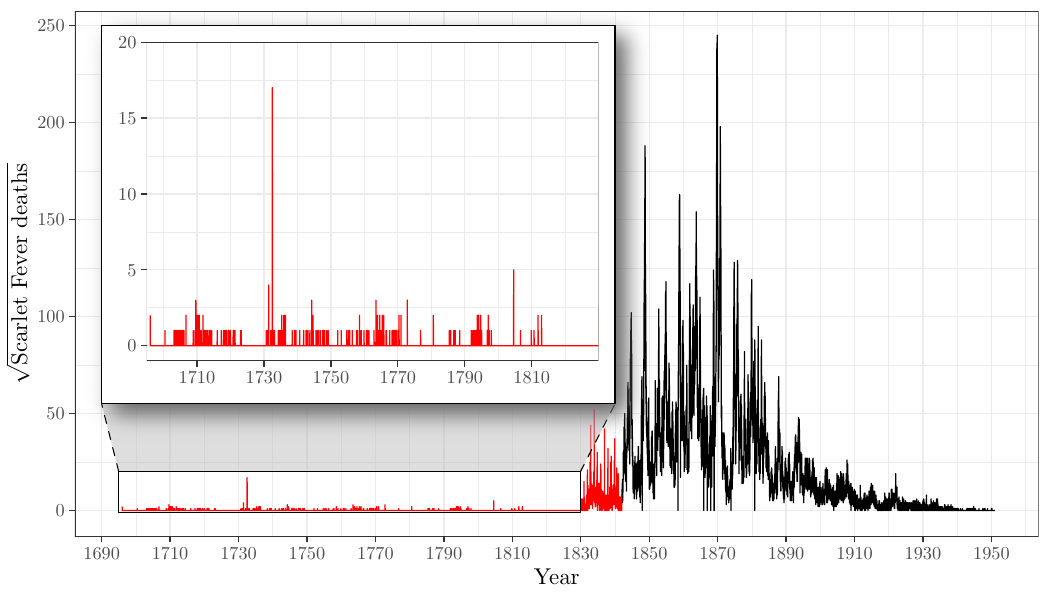} \caption{Observed scarlet fever deaths in the LBoM (red, before 1842) and RGWR (black, 1842 and onward).  The first recorded scarlet fever death in the LBoM occurred on the week of January 7th, 1696.}\label{fig:LBoM}
\end{figure}

\begin{figure}[p]
    \includegraphics[width=\maxwidth]{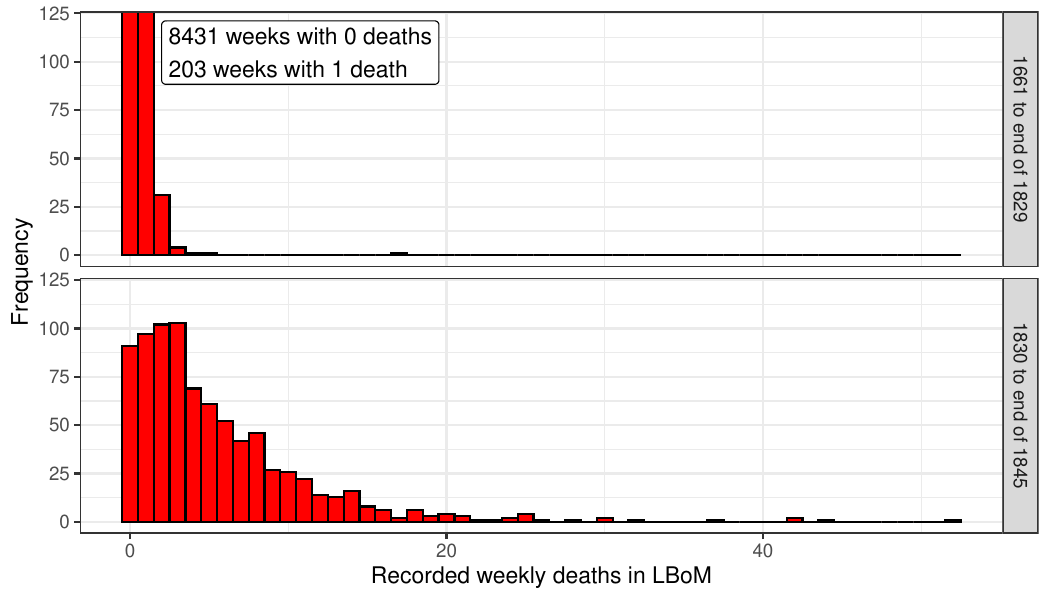} \caption{Histogram of the counts of weekly deaths recorded in the LBoM, separated into the periods from 1661 to 1829 (during which period $0$ deaths were recorded after 1813 and the highest number of recorded weekly deaths was $17$), and from 1830 to 1845.}\label{fig:LBoMhist}
\end{figure}

Weekly mortality from SF in the city of London across individual parish registers has been recorded since
the late 17th century by the Company of Parish Clerks (\cref{fig:LBoM}).  The first SF deaths were
recorded in the London Bills of Mortality (LBoM) for the week of 7
January 1696.  The reported weekly SF mortality until the week of 8
December 1812 was typically zero or one.  Creighton
\citep[p.\,719]{Crei1894} notes that SF was mentioned in the writings
of a number of physicians in the late 18th and early 19th centuries,
but was frequently misclassified as measles or fever in the London
Bills of Mortality (LBoM).  It was only in 1830 that the parish clerks
began to properly record SF in the LBoM, under the categories
``scarlet fever'' and ``scarlatina''.  In these categories, there were
303 deaths recorded in the LBoM from 1696 to 1812, none from 1813 to
1829, and 4,585 from 1830 to 1845 \citep[p.\,719]{Crei1894}.  (See the
histograms in \cref{fig:LBoMhist}.)
The LBoM are known to have several data quality issues, such as only reporting deaths of those given Anglican burials, not including new regions of London after several boundary expansions, and, most importantly, the progressive collapse of the parish registration system, where certain parishes stopped submitting returns after 1830 \citep[p.\,3]{KrylEarn20}, leading to underreporting of deaths.

\subsection{SF in the Registrar General's Weekly Returns, 1842--1950}

With the creation of the Registrar General's Office in 1836, and a new
civil registration system in 1837, the parish registration system
that gave rise to the LBoM was superseded by the Registrar General's
Weekly Returns (RGWR) in 1842.  The RGWR captured registered deaths
beyond Anglican burials and covered a larger, though still evolving,
geographic definition of London; several districts were added during
the 1840s, while some suburban localities remained outside its limits
\citep[p.\,5]{KrylEarn20}.
For our purposes of modelling scarlet fever, we use only the data from the RGWR, as the vast majority of scarlet fever deaths reported in the LBoM occur between 1830 and 1845, during the gradual collapse of the parish registration system.
Consequently, we have no way of knowing to what extent the data are affected by missing parish returns after 1830, making reliable consolidation of the LBoM and RGWR SF series impossible.
A plot of the LBoM and RGWR data is shown in \cref{fig:LBoM}, from which we can see that the LBoM captured significantly fewer deaths.
As such, our analysis from hereon will focus only on data from the RGWR.
The digitized RGWR contained 37 rows with missing SF data and omitted
35 additional weekly rows.  The longest consecutive gap was the nine
weeks from 3 March to 11 May 1929.  We inserted the omitted weeks,
corrected minor date-entry errors, and imputed all missing observations
by linear interpolation.

SF mortality remained high until approximately 1880, after which it
declined substantially \citetext{\citealp[p.\,719]{Crei1894};
  \citealp[p.\,112]{Kass71}; \citealp[p.\,1113]{Floy94}}, especially
after the introduction of new treatments (sulfonamide in 1935 and
penicillin in 1945 \citetext{\citealp[p.\,1113]{Floy94};
  \citealp[p.\,112]{Kass71}}).

The RGWR also began reporting new weekly cases for notifiable infectious diseases in 1901, with the first week of case reporting being the week starting 29 December 1901.
Although much less lethal at this point in time, SF was one of these notifiable diseases, remaining highly prevalent despite very low mortality.
Scans of the RGWR from 1840 to 1967 are publicly available from the
Wellcome Collection through the Internet Archive \citep{Wellcome}.
ChatGPT and OpenAI Codex assisted with transcription of the weekly SF
case counts used here.  A sample of the transcribed entries was
checked manually against the original scans.

\subsection{SF mortality data quality assurance}

\begin{figure}[p]
\includegraphics[width=\maxwidth]{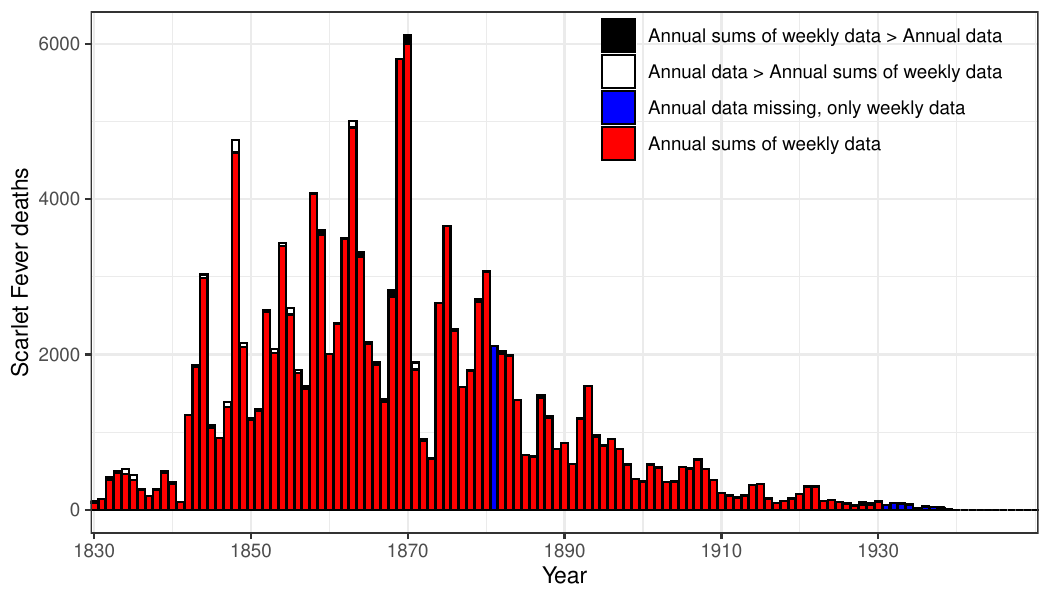} \caption{Consistency check comparing the annual sums of weekly scarlet fever mortality to annual mortality data.}\label{fig:Annual_plot}
\end{figure}

We can verify the consistency of the SF mortality by comparing the
annual sums of weekly deaths from the LBoM and RGWR to published
annual mortality records:
\begin{itemize}
  \item From 1830 to 1842, annual data are from the Weekly Return of Births and Deaths and the Weekly Bills of Mortality, as the parish clerks also collected annual returns.  
  \item From 1842 to 1852, annual data are sourced from ``Births Deaths and Causes in London with Meteorological Observations For the Thirteen Years 1840-1852.''
  \item From 1853 to 1930 annual data are from the RGWR, which similarly published annual summaries.
\end{itemize}
A comparison of annual data from these sources is presented in
\cref{fig:Annual_plot}.  Agreement between the data sets when annual data are available is close, as
evidenced by the low heights of the black and white disagreement bars
stacked on the red bars.

\subsection{Vital statistics}

The RGWR reported weekly births beginning the week ending 2 January
1842.  We have previously digitized these birth data \cite{KrylEarn20}
until the week ending 3 January 1931, with the exception of one full
missing year 1881 and five missing weeks\footnote{Weeks starting on
26 February 1842, 14 May 1842, 17 June 1843, 21 December 1844, and 25
December 1847.}.  We imputed the missing data using
linear interpolation.  We use only the trend of the birth rates, so
the effects of missing birth data are negligible.  We estimated the
trend using empirical-mode decomposition \citep{Huang+71,KrylEarn20},
wherein zero-mean periodic waves in the data (called intrinsic-mode
functions) are extracted, beginning with lower periods.  This process
is repeated until what remains of the time series is a close
approximation of the trend, which is smooth and has no more
oscillations.
For births after 1931, we used the same AI-assisted transcription
procedure on the digitized RGWR scans.  As a validation check, the
AI-assisted weekly birth counts agreed exactly with independently
hand-digitized values for every week of 1930.
Likewise, all-cause mortality (ACM) data were reported and digitized starting in 1842 as well, but with weekly records having been hand digitized until 1950.
Like the birth data, we only use the trend of the ACM data.
Population data for London were also collected by
census once every decade beginning in 1801 \citep{Demographia}.  We
estimated the weekly population by performing linear interpolation
between decades, following~\cite{KrylEarn20}.

\begin{figure}[p]
  \centering
  \includegraphics[height=0.72\textheight,keepaspectratio]{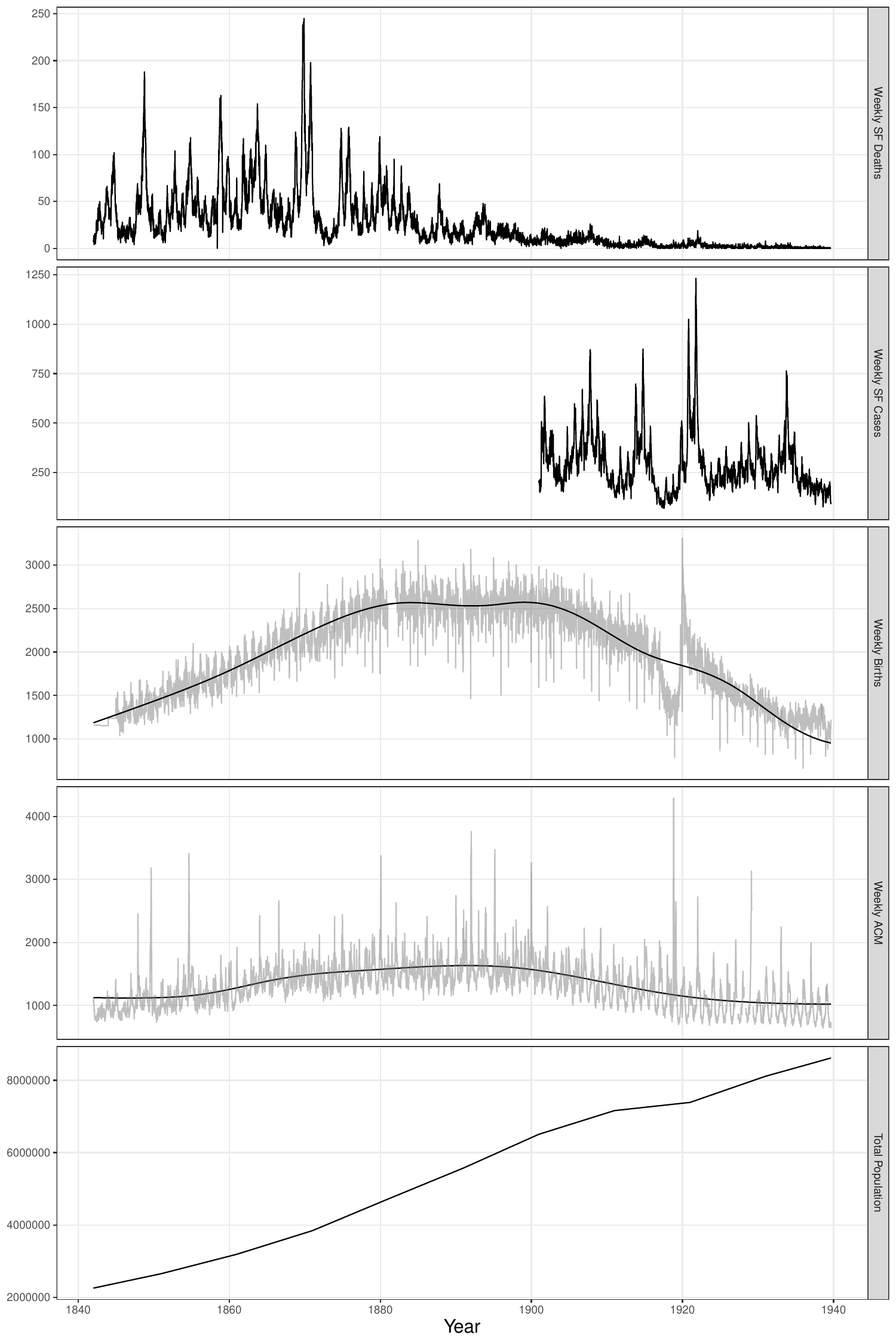} \caption{(From top to bottom) Weekly scarlet fever deaths, AI-assisted transcription of weekly scarlet fever cases, raw counts (gray) and trend (black) of weekly births, raw counts (gray) and trend (black) of weekly ACM, and interpolated total London population.
The weekly birth trend was estimated using empirical-mode decomposition \citep{Huang+71}.
}\label{fig:BirthSeries}
\end{figure}

\begin{figure}[p]
  \includegraphics[width=\textwidth]{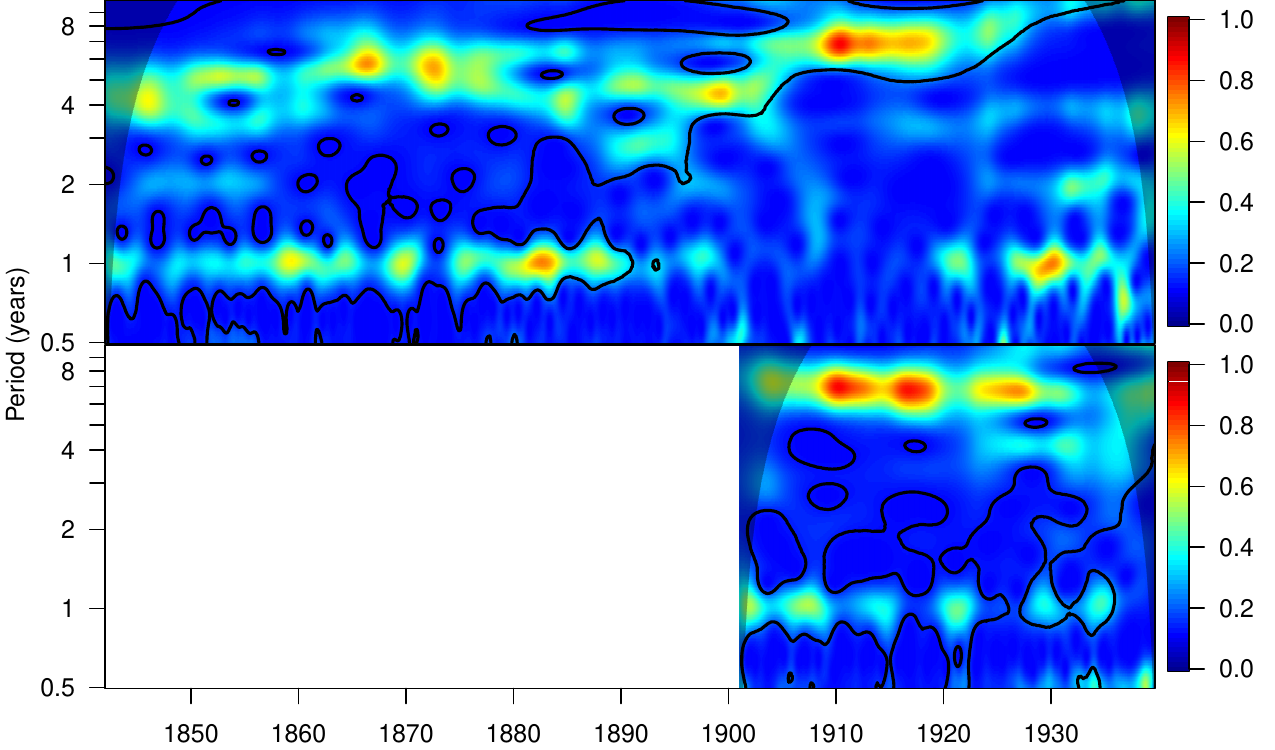} \caption{Wavelet spectra of the weekly SF mortality (top) and cases (bottom).  Hotter regions represent stronger presence of the period at the given time point.  Wavelet transformations were computed using the \texttt{WaveletComp} R package \cite{RoesSchm18}.}\label{fig:Wavelets}
\end{figure}

Although the RGWR data are available until 1967, we end our analysis
with the week ending 2 September 1939, immediately before Great
Britain declared war on Germany during the Second World War
\cite{IWM}.
During this time, 1.5 million Londoners evacuated to rural locations, many of whom were children \cite{IWMevac}.
This large population shift would have substantially affected SF
dynamics and complicated the demographic assumptions of our model.
The data used in our analysis are summarized in \cref{fig:BirthSeries}.
We also summarize the changing dynamics of SF over time via the wavelet spectrum of both the mortality and case data in \cref{fig:Wavelets}.
Despite the sensitivity of the low mortality counts to noise, reflected
in the absence of a clear annual cycle after approximately 1890, the
wavelet spectra of weekly deaths and cases both show an approximately
eight-year periodicity from 1905 to 1930.

\section{Description of the Model}

Following Refs.\,\citep{HempEarn15, KrylEarn13},
we write the SIR model as a system of differential equations, 
\begin{subequations}\label{eq:SIR}
  \begin{linenomath*}
    \begin{align}
      \frac{{\rm d}S}{{\rm d}t} &= \nu N_0 - \beta SI - \mu S, \label{eq:SIR;S}\\
      \frac{{\rm d}I}{{\rm d}t} &= \beta SI - \gamma I - \mu I, \label{eq:SIR;I}\\
      \frac{{\rm d}R}{{\rm d}t} &= \gamma I - \mu R. \label{eq:SIR;R}
    \end{align}
  \end{linenomath*}
\end{subequations}
The state variables, $S$, $I$, and $R$ represent the numbers of
susceptible, infected, and recovered individuals.  Since $R$ appears
only in \cref{eq:SIR;R}, we may remove it and consider the system in
the two variables $S$ and $I$.  New susceptibles enter the population
via the $\nu N_0$ term, where $\nu$ is the birth rate relative to the
population ($N_0$) at some anchor time ($t_0$).  We introduce newborns
directly into the susceptible class.  This ignores temporary protection
from maternal antibodies during infancy \citep{KatzMore92}, an
approximation that we consider further in the Discussion.  Susceptibles flow into the infected class
due to the transmission term, $\beta SI$, where $\beta$ is the
transmission rate.  Infecteds recover at rate $\gamma$, where
$\gamma^{-1}$ is the mean infectious period.  The rate $\mu$ captures
death by non-SF causes and net population change due to migration.
For simplicity, we
assume that recovery confers lifelong immunity, whereas in practice
full immunity to SF tends to require repeated infection
\citep[p.\,8]{DickDick34}.  We consider the limitations of this
assumption in our Discussion.

With this model \eqref{eq:SIR}, we can interpret transmission rates
more easily using $\R_0$, the basic reproduction number, which
represents the expected number of susceptibles to be infected directly
by a single infected in an entirely susceptible population
\citep{AndeMay91}.  $\R_0$ is related to the parameters in
\cref{eq:SIR} via \citep{KrylEarn13}
\begin{equation}\label{eq:R0}
  \R_0 = \frac{\nu N_0}{\mu} \frac{\beta}{\gamma + \mu}.
\end{equation}

The SF transmission rate varies seasonally.  Case counts are higher
when schools are open \citep{Brow1920}.  The
detailed pattern of seasonality of transmission is unknown, so we
write the transmission rate
\begin{equation}\label{eq:force}
  \beta(t) = \beta_0 (1 + \alpha g(t)),
\end{equation}
where $\beta_0$ is the mean transmission rate, $\alpha$ is the
amplitude of seasonal forcing, and $g(t)$ is the seasonal forcing
function, \ie a periodic function with period $1$ year, mean $0$, and
$\max(|g(t)|) = 1$ \citep{PapsEarn19, HempEarn15, Earn09}.
\cref{eq:R0} is still valid with seasonal forcing if we replace
$\beta$ with $\beta_0$ \cite{MaMa06}.

The true forcing function $g(t)$ can be estimated from disease
incidence or mortality data \cite{Jaga+20}.  Previous work has shown
that the dynamics induced by any such forcing function $g(t)$ are
qualitatively equivalent to the dynamics induced by sinusoidal forcing
($g(t)=\cos{(2\pi t)}$), provided we adjust the amplitude ($\alpha$)
of seasonality appropriately (via matching the position of a
period-doubling bifurcation \cite{PapsEarn19}).  Consequently, we can
use the model \eqref{eq:SIR} with sinusoidal seasonal forcing to
predict transitions in the periodic structure of epidemic patterns, as
the reproduction number and sinusoidal forcing amplitude change
\cite{Earn+00,BaucEarn03a,Earn09,KrylEarn13,HempEarn15,PapsEarn19}.

We use the SIR model \eqref{eq:SIR} rather than the more realistic
SEIR model, which accounts for the delay between initial infection and
becoming infectious, because previous work \cite{KrylEarn13} has shown
that the SEIR model yields periodicities that match those of the SIR
model with the same mean generation interval (time from being infected
to infecting another individual \cite{ChamDush15}), so the simpler
model can be used for analysis.  The mean generation interval in the
SEIR model is the sum of the mean latent period and the mean
infectious period ($\sim1$ day and $\sim14$ days for SF, respectively;
see \cref{tab:Params}).  In the SIR model, the mean generation
interval is equal to the mean infectious period, so we set
$\gamma^{-1}=15$ days.

Over the span of a few years, the model as described above with
constant parameter values is adequate, but over decades, birth rates
changed in a known manner (so $\nu=\nu(t)$, which demographic data
allow us to make precise).  In addition, both the mean transmission
rate $\beta_0$ (and hence $\R_0$), and amplitude of seasonality
$\alpha$, changed as population density and mixing patterns changed
over time.  Thus we must estimate $\R_0(t)$ and $\alpha(t)$, which we
explain in the following sections.

\section{Estimating $\R_0(t)$ and $\alpha(t)$ with \texttt{macpan2}}

To estimate $\R_0$ and $\alpha$, we must estimate the function
$\beta(t)$, giving us the shape and magnitude of the transmission rate
every year.  To do this based on the observed mortality and case data, we
revise the SIR model \labelcref{eq:SIR} to include mortality, $M$,
explicitly.  The weekly flow into $M$ gives expected deaths, while the
weekly flow from $S$ to $I$ through transmission gives expected cases.
We fit these two observation series jointly by maximum likelihood:
mortality contributes throughout 1842--1939 and cases additionally
contribute from 1901 onward.

The revised model equations are
\begin{subequations}\label{eq:SIRmac}
	\begin{linenomath*}
		\begin{align}
			\frac{{\rm d}S}{{\rm d}t} &= B - \beta \frac{SI}{N} - D \frac{S}{N}, \label{eq:SIRmacS}\\
			\frac{{\rm d}I}{{\rm d}t} &= \beta \frac{SI}{N} - \gamma I - D \frac{I}{N}, \label{eq:SIRmacI}\\
		\frac{{\rm d}R}{{\rm d}t} &= (1 - \CFP) \cdot \gamma \cdot I - D \frac{R}{N}, \label{eq:SIRmacR}\\
		\frac{{\rm d}M}{{\rm d}t} &= \gamma \cdot \CFP \cdot I. \label{eq:SIRmacM} 
		\end{align}
	\end{linenomath*}
\end{subequations}
Here, $N$ is the time-varying population size, $B$ is the weekly
number of births, and $D$ is the weekly net demographic outflow;
$\CFP$ is the time-varying case-fatality proportion, so the rate
of SF-induced death is $\gamma\cdot\CFP\cdot I$.  Population accounting
over week $k-1$ gives
\begin{subequations}\label{eq:demographic-removal}
\begin{align}
  N_k-N_{k-1} &\;=\; B_{k-1}-D_k-(M_k-M_{k-1}), \label{eq:demographic-removal;a}
\intertext{or, equivalently,}
  D_k &\;=\; B_{k-1}-(N_k-N_{k-1})-(M_k-M_{k-1}). \label{eq:demographic-removal;b}
\end{align}
\end{subequations}
Thus $D_k$ is births minus population growth and SF deaths over that week.  It
represents deaths from all causes aside from SF (which are accounted for by the $M$ state) plus emigration minus immigration,
and can be negative if net immigration exceeds deaths.  This is an
approximate quantity because $N_k$ comes from the interpolated
population series and $B_k$ from the estimated birth trend.

One would typically assume the CFP to remain constant for a disease.
However, given that we have a relatively long time series spanning
several decades, and that it is hypothesized that by the 1890s the
dominant strain of SF had changed to one that was much milder
\citep[p.\,65]{Hard93}, we chose to model the CFP as a logistic curve,
with CFP being larger at the beginning of the time series and smaller
at the end.  More details of this CFP estimation can be found in
\cref{appendix:InitialValues}.

Some parameters are known \emph{a priori} or estimated from data other
than our focal scarlet fever time series; these parameter estimates,
which we took as given, are summarized in \cref{table:pars}.  To
estimate $\beta(t)$, we then used the \texttt{macpan2} R package
\citep{Walk+25} to fit the model \labelcref{eq:SIRmac} to the scarlet
fever time series.
\texttt{macpan2} allows us to specify a differential equation model of
infectious disease transmission, and then estimate model parameters by
minimizing an objective function, in our case the negative log-likelihood of
the observed SF mortality and incidence series, assuming observation
errors are negative-binomially distributed.
By default, \texttt{macpan2} uses Euler's method for ODEs, but since
our time steps are large (weekly), we used the fourth order
Runge-Kutta method to ensure the \texttt{macpan2} solution remains
close to the true differential equation's behaviour.

The fitting model uses incidence $\beta SI/N$ and allows population
size and transmission to change over the historical record.  For
transition analysis, we consider temporal segments over which the
trends in births, population size, and mean transmission change little.
Within each segment, we approximate these trends as fixed while retaining
seasonal variation in transmission.  If $t$ denotes a representative
time in a segment, the fitted mean incidence term is approximated by
the mass-action term $\beta_{\rm MA}(t)SI$, where
$\beta_{\rm MA}(t)=\beta_{\rm trend}(t)/N(t)$.  The quantities on the
right-hand side are evaluated at this representative time and treated
as constant within the segment.  The birth input is the observed trend
in weekly births, $B(t)$, likewise treated as constant within the
segment rather than made proportional to the model's changing population
size.  Changes in $B(t)$ between segments can therefore induce dynamical
transitions \citep{Earn+00,KrylEarn13}.

We set $\mu(t)=A(t)/N(t)$, where $A(t)$ is the trend in weekly
all-cause mortality.  In the absence of infection, births balance
mortality at $S=B(t)/\mu(t)$.  Substituting this value and
$\beta_{\rm MA}(t)$ into \cref{eq:R0} gives
\begin{equation}\label{eq:R0-timevarying}
  \R_0(t)=
  \frac{\beta_{\rm trend}(t) B(t)}{A(t)}
  \left(\gamma+\frac{A(t)}{N(t)}\right)^{-1}.
\end{equation}
We use $A$ rather than the fitted model's net population outflow $D$
because $D$ includes migration and can be negative.  The resulting
$\R_0(t)$ locates the historical parameter trajectory in the
mass-action transition analysis.

\subsection{Modelling time-varying parameters}

In the model \labelcref{eq:SIRmac}, the transmission rate $\beta$ is a
time-varying function given by \cref{eq:force}, which contains three
potentially time-varying components: the magnitude ($\beta_0$), the
shape of the forcing function ($g(t)$), and the amplitude of forcing
($\alpha$).  We need to estimate these three components in order to
capture and identify temporally where the dynamical transitions in the
scarlet fever series occur.

During fitting we do not estimate the normalized forcing function
$g(t)$ and its amplitude $\alpha$ separately.  Instead, we represent
the amplitude-scaled seasonal component by a three-term Fourier series
whose coefficients vary smoothly over time.  Adding further Fourier
terms did not change the results qualitatively.  We denote the fitted
mean transmission rate by $\beta_{\rm trend}(t)$ and the fitted,
amplitude-scaled seasonal component by $g_{\rm mac}(t)$.

The mean transmission rate and the Fourier coefficients are expected to be
smooth functions of time.  To estimate these functions, we use a set
of $m$ Gaussians evenly centred across our time series as Radial Basis
Functions (RBFs)\footnote{An RBF is a function with one input vector
whose value is solely dependent on its input's Euclidean distance from
some centre $\mathbf{x}_i$, or equivalently, a function of the form
$f(||\mathbf{x}-\mathbf{x}_i||)$.  These functions are typically
easily computed and form a basis with which we can approximate a
function as a linear combination of the RBFs \citep{Light92}.  For
our modelling purposes, we have $\mathbf{x}$ as a scalar.} and
respective weights to be used across $n$ time steps.  We had $n =
6055$ time steps (weeks), and found that $m=64$ RBFs overfit the
data, as shown in the top panel of \cref{fig:overfit}.  To account for
this overfitting, we used a penalized likelihood method, where the RBF
weights for the Fourier coefficients $\mathbf{a}_j, \mathbf{b}_j$ for
$j = 1,2,3$ and for $\beta$'s magnitude, $\mathbf{c}$ were modelled
under a normal distribution with mean $0$ and standard deviations of
$\sigma_{\rm f}$ and $\sigma_{\beta}$ respectively (penalization
reduced the fitted weights' amplitudes).  The likelihoods of these
weights were then included in a negative log-likelihood function,
alongside the negative log-likelihoods of the observed mortality and
case data under negative binomial distributions with means given by
the corresponding simulated flows and fitted dispersion parameters
$\phi_M$ and $\phi_C$.  \texttt{macpan2}
minimizes this likelihood function using gradient methods, treating
all parameters as fixed effects by default.  Given that we have no
mechanistic interpretation of the RBF weights, they should in
principle be treated as random effects, however we found that using
fixed effects was much more computationally efficient and produced
nearly identical results to treating the weights as random effects.
These weights represent changes in their respective coefficients over
time, with later indexed weights more significantly impacting
$\beta(t)$ for larger $t$.  Each RBF has maximum value $1$, with the
$i$th Gaussian being centred at $x_i = 1 + \frac{(n-1) \cdot
  (i-1)}{m-1}$ with standard deviation $n/m$, or equivalently,
\begin{equation}\label{eq:RBF}
	f_i(x) = \exp\left(\frac{-\left(x-x_i\right)^2}{2\left(\frac{n}{m}\right)^2}\right).
\end{equation}
We can then define the vector of RBFs
\begin{equation}\label{eq:vecRBF}
	\mathbf{F}(t) = \begin{bmatrix} f_1(t) \\ f_2(t) \\ \vdots \\ f_{m-1}(t) \\ f_m(t) \end{bmatrix}.
\end{equation}
To estimate the magnitude and trend of $\beta$ over time as a continuous function, we use
\begin{subequations}\label{eq:macbeta}
	\begin{linenomath*}
		\begin{align}
			\beta_{\rm trend}(t) &= \exp\left(b_0 + \mathbf{F}(t) \cdot \mathbf{c}\right),\\
      g_{\rm mac}(t) &= \mathbf{F}(t) \cdot \sum_{j=1}^3 \Big(\mathbf{a}_j \cos\frac{2\pi j t}{T} + \mathbf{b}_j \sin\frac{2\pi j t}{T}\Big), \\
      \beta_{\rm mac}(t) &= \beta_{\rm trend}(t) \left(1 + g_{\rm mac}(t)\right),
		\end{align}
	\end{linenomath*}
\end{subequations}
where $t$ represents the time step, in units of weeks, and $T = 365.25/7$ weeks is the forcing period.  The exponential link ensures that
$\beta_{\rm trend}(t)$ remains positive.  Because $g_{\rm mac}(t)$
already contains the seasonal amplitude, it is not normalized in the
manner of $g(t)$ in \cref{eq:force}.  For transition analysis we obtain
an equivalent sinusoidal amplitude separately for each time interval
by matching a period-doubling bifurcation, as described below.
We show the fitted values in our \texttt{macpan2} model in \cref{tab:ParamTable}, and display the fitted mortality series in \cref{fig:macpanfit}.
To account for inaccuracies in fitting transmission rate at the boundaries of our time series, where edge effects cause $\beta(t)$ to converge values with extreme behaviour, we duplicated the first $479$ and last $508$ weeks of our time series, thus yielding $n = 6055$ time steps from $5068$ weeks of scarlet fever data.
We truncated these replicated years before creating our plots.

\section{Comparison of \texttt{macpan2} results with \texttt{fastbeta}}

\texttt{fastbeta} is an R package that estimates $\beta(t)$ in SIR and SEIR models at discrete time points using disease incidence, birth, and population removal time series, alongside the mean generation interval and an initial estimate of each disease state in the population \citep{Jaga25, Jaga+20}.
When incidence data are unavailable, disease mortality data can be given alongside the CFP to deconvolve incidence from mortality using a method based on Richardson--Lucy deconvolution \cite{Gold+09}.

\begin{figure}[p]
	\includegraphics[width=\maxwidth]{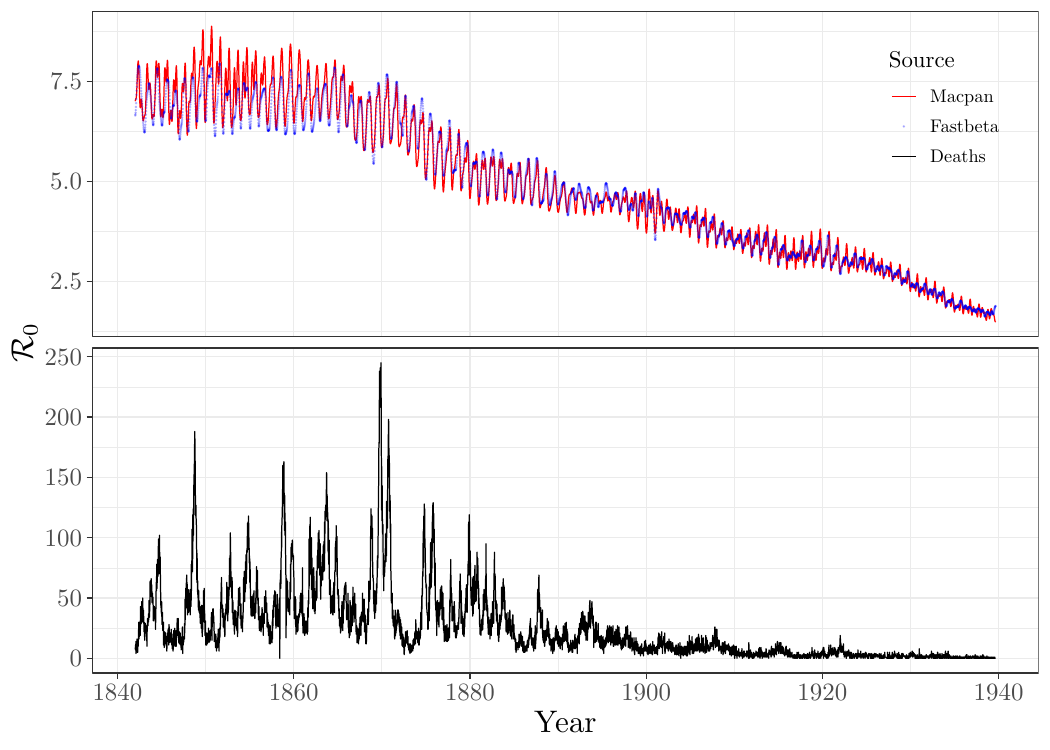} \caption{The blue points represent estimates of $\beta_t$ from \texttt{fastbeta} plugged into the $\R_0$ equation \eqref{eq:R0-timevarying}.
  As recommended in \cite{Jaga+20}, we smoothed the \texttt{fastbeta} $\beta_t$ estimates by taking the loess curve with a span of the nearest $53$ points before plotting.
For comparison, we display the fitted $\R_0$ curve from \texttt{macpan2} in red.
Note that both methods used in this figure use the time series with extra weeks of padding at their beginning and end.
}\label{fig:Betas}
\end{figure}

\begin{landscape}
\begin{figure}[p]
  \centering
	\includegraphics[width = 1.4\maxwidth]{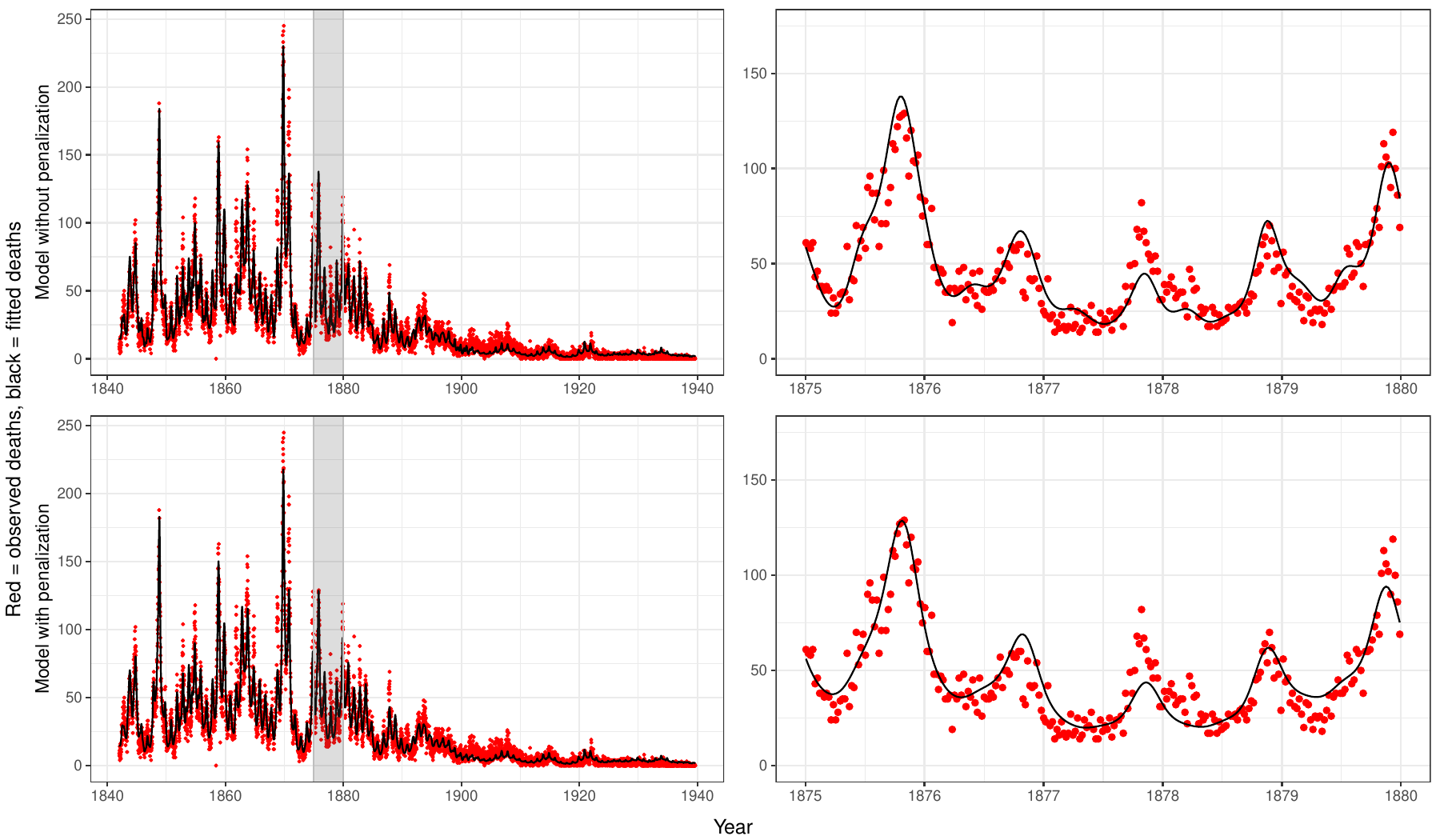}\caption{
    Comparison of non-penalized \texttt{macpan2} fit and penalized \texttt{macpan2} fit (left), with 1875 to 1880 highlighted as an example of the local differences in fitting (right).
In the non-penalized model, the solution attempts to capture sub-annual patterns in the data which appear to be noise.
The penalized model appears to be less influenced by this noise, with signals only appearing at seasonal peaks and troughs of the data.}\label{fig:overfit}
\end{figure}
\end{landscape}

\begin{figure}[p]
  \includegraphics[width=\textwidth]{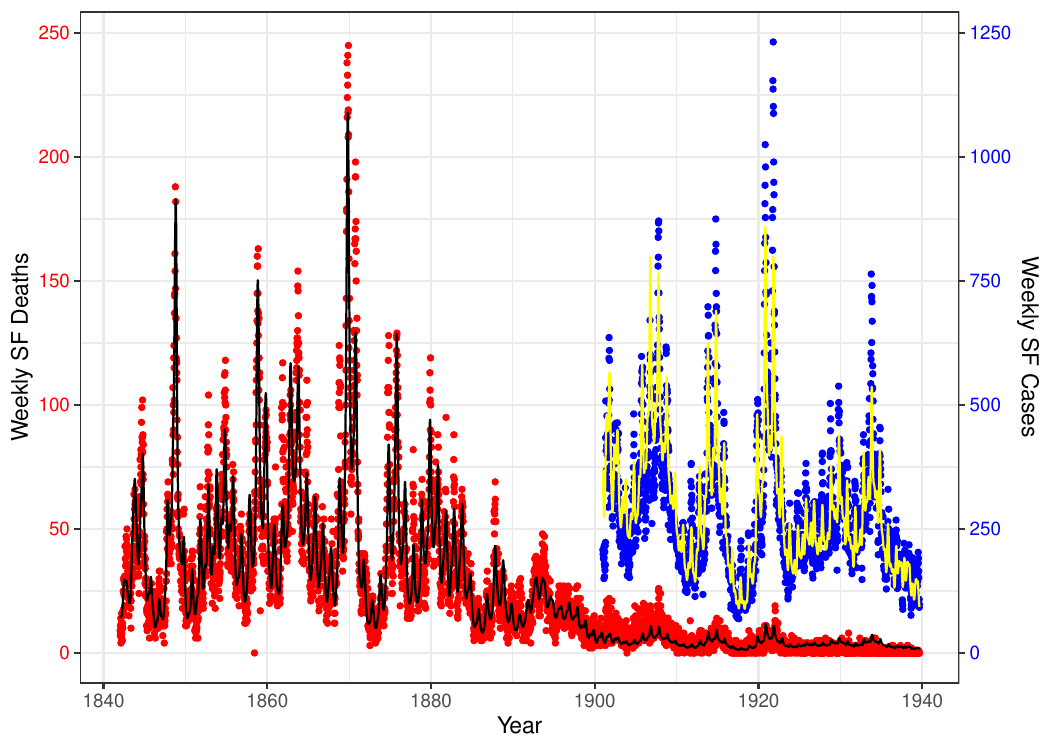} \caption{\texttt{macpan2} fitted deaths (black line) and observed deaths (red line).}\label{fig:macpanfit}
\end{figure}

As a robustness check on the fitted $\beta(t)$ curve obtained from
\texttt{macpan2}, we used \texttt{fastbeta} to estimate the discrete-time
transmission rate $\beta_t$ and compared the resulting trajectories.
For our comparison, we used \texttt{fastbeta} to estimate $\beta_t$ for our SIR model using the fitted CFP and population removal rate time series and fitted initial values of $S, I, R$ from \texttt{macpan2}.
To get estimates from \texttt{fastbeta}, we used the observed weekly SF mortality data prior to 1901, deconvolving it using the fitted CFP.
For the deconvolution, we assumed the delay between infection and death was exponentially distributed with mean equal to the mean generation interval.
After 1901, the mortality data were far more susceptible to random noise, so we used the state \texttt{fastbeta} estimated for the last week of 1900 and input the observed weekly SF case data.
Comparing the $\beta(t)$ trajectories obtained from the same data, we see in \cref{fig:Betas} that the two packages yield nearly identical values for $\R_0$, apart from roughly 1850--1870, when the \texttt{macpan2} estimate is consistently greater than the \texttt{fastbeta} estimate.
The phase and amplitudes of the various peaks and troughs also appear to line up consistently.

\section{Transition analysis}

With our model defined, we seek to explain transitions in the scarlet
fever mortality series in terms of changes in the parameters of our
SIR model.  We identify the periodicities that the model predicts at
each point in parameter space through simulation.  From the fitted
\texttt{macpan2} model we then derive a parameter trajectory and ask
whether the dynamical transitions predicted along it agree with the
observed dynamics.
In all our simulations of the model, we simplify the model in \cref{eq:SIR} by removing the $R$ compartment, keeping the overall population size constant by setting $\mu = \nu = \SI{0.02}{yr^{-1}}$, and using proportions rather than population counts by setting $N_0 = 1$.
Short-term changes in both birth and death rates are relatively small, so we can consider them constant on typical epidemic timescales even though they vary over longer periods.
This allows us to make conclusions about the long-term behaviour predicted by a given transmission rate.

\begin{landscape}
\begin{figure}[htb!]
  \includegraphics[width=\linewidth]{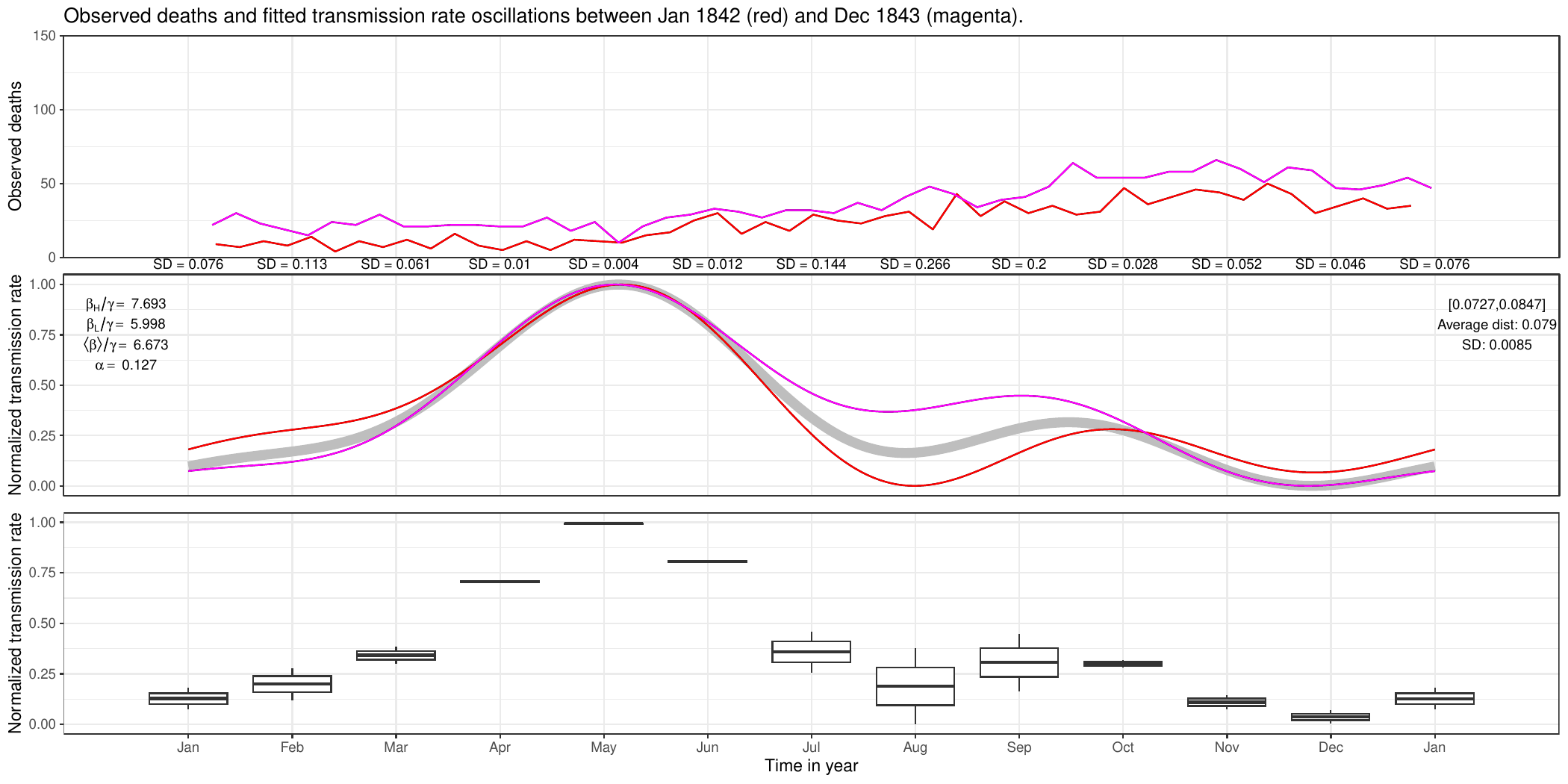} \caption{Top: Weekly scarlet fever deaths for the years of 1842 and 1843 overlaid on each other.
  Middle: Transmission rates across these years with various statistics shown on the graph.
Average transmission rate shape shown in thick grey, obtained by averaging the values of the Fourier weights at the beginning of each of these years, and then using the average weights for calculating the shape component of $\beta_{\rm mac}(t)$ in \cref{eq:macbeta}.
Note that in this procedure, we assume the transmission rate shapes are not changing dramatically within each year, as we only take the transmission rate at the beginning of each year.
Between the top and middle plot are standard deviations of the yearly transmission rates at the beginning of each month.
Bottom: Boxplots of the yearly transmission rates at the beginning of each month.
We evaluated plots like this for each part of our partitioning of the time series to determine our breakpoints.
With our partitioning of the time series, the transmission rates have low variance within the partition, as the boxplots do not have major vertical spread.}\label{fig:boxplots}
\end{figure}
\end{landscape}

A natural way to begin is to examine how periodicities in the mortality series change as one parameter is varied with the other parameters held fixed.
We know that part of what drives these periods in the model is our transmission rate, $\beta(t)$, which from \cref{eq:macbeta}, is controlled by its shape and magnitude.
For a fixed shape, simulating the results of the model across many magnitudes of $\beta(t)$ is simple, as we may simply run the model until it converges, and note the period of the attractor, repeating for many initial conditions and increments of $\R_0$, which we use to represent $\beta_0$ through \cref{eq:R0} as it is much easier to interpret.
However, $\beta(t)$ has a constantly changing shape function over time, to represent different scarlet fever epidemics, which makes it impossible to make conclusions about the long-term behaviour of the model for the infinite possible shapes of $\beta(t)$.
To make use of the shape of $\beta(t)$ fitted by \texttt{macpan2}, we first partition the time series into intervals within which its seasonal shape does not vary substantially.
We selected the breakpoints by visual inspection.  The distributions shown in \cref{fig:boxplots} indicate that the mean seasonal shape within each interval provides a reasonable approximation to the fitted shapes for its constituent years.
This shape is a major determinant in the dynamics predicted by our model, hence it is vital that we keep as much information from this shape as we can while reducing it to a more manageable quantifier.
One method to do this was demonstrated in~\citet{PapsEarn19}, which found that, if we fix the shape of a transmission rate, we can obtain an $\alpha$ such that the sinusoidal forcing function (when $g(t) = \cos(2\pi t)$) yields invariant dynamics across $\R_0$ values.
This gives us a method to capture the dynamics of a myriad of possible shape functions by only varying $\alpha$ while keeping the shape fixed.

\begin{figure}[htb!]
  \includegraphics[width = \linewidth]{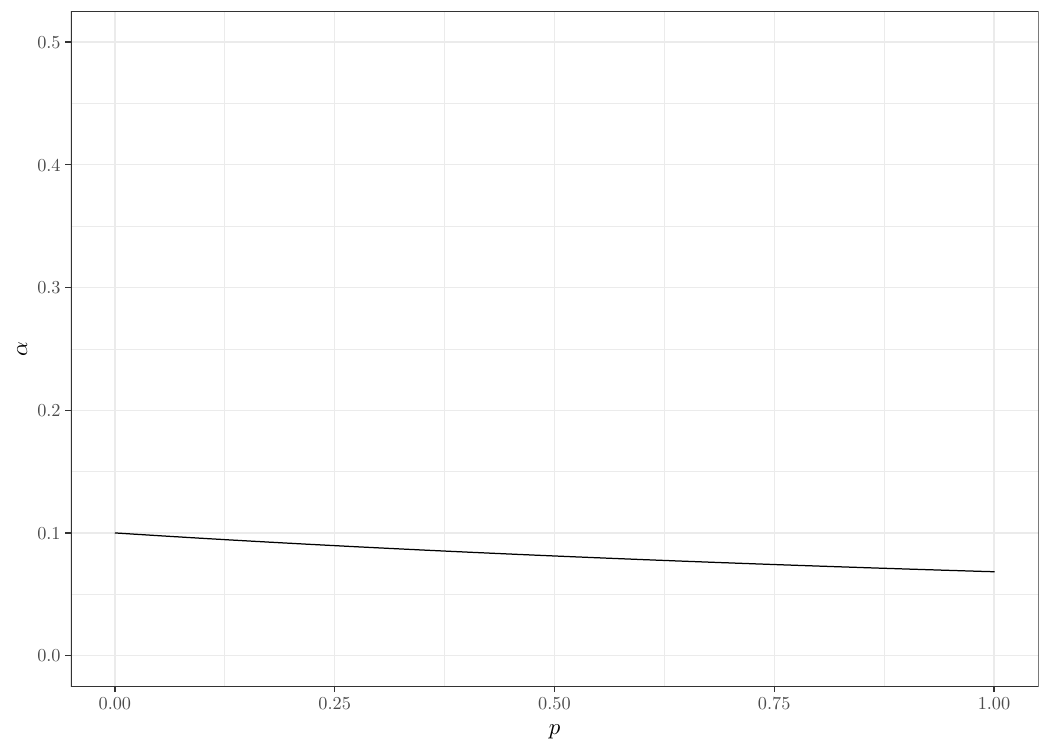} \caption{Two-parameter bifurcation diagram on amplitude ($\alpha$) and shape parameter ($p$), where $p = 0$ represents the average \texttt{macpan2} forcing function from the beginning of 1842 to the end of 1844 and $p = 1$ represents sinusoidal forcing.
  We follow the period doubling bifurcation at $\R_0 \approx 19.46$ in \cref{fig:AttTran}.}\label{fig:alpha_cont}
\end{figure}

The equivalent sinusoidal amplitude for a given transmission-rate shape can be found using \texttt{XPPAUT} \citep{Erme02}.
The supplementary materials of~\cite{PapsEarn19} describe this procedure in greater detail, but to summarize our approach:
\begin{enumerate}
  \item We define the SIR model, and provide $\beta(t)$ as the function given below in \cref{eq:betap}
    \begin{equation}
      \beta(t) = \beta_0 \left( 1 + \alpha \cdot ((1-p)\cdot g_{\rm mac}(t) + p \cdot \cos(2\pi (t - t_0)) \right),\label{eq:betap}
    \end{equation}
    where $g_{\rm mac}(t)$ is the average transmission rate function from a given part of the \texttt{macpan2} transmission rate shapes partition, $t_0$ is a phase parameter, representing the time of year when $g_{\rm mac}(t)$ attains its maximum (so that both forcing functions attain their maximum value at the same time, and are more similar in shape) and $p$ represents a shape parameter to be varied on $[0, 1]$, with $p = 0$ representing \texttt{macpan2} forcing and $p = 1$ representing sinusoidal forcing.  
    To simplify interpretation, we write $\beta_0$ as a function of $\R_0$, and use $\R_0$ as a variable.
  \item We simulate the model under \texttt{macpan2} forcing by setting $p = 0$ and $\alpha = 1$ (as $g_{\rm mac}(t)$ is already scaled by $\alpha$).  
    Our goal is to converge to an annual attractor, which is dependent on $\R_0$, $S_0$, $I_0$.
    This can be done through trial and error, such as by using initial conditions $S_0 = 0$, $I_0 = 0.001$ and trying various values of $\R_0$ near $15$.
    Basins of attraction for the annual attractor tend to be large for the shape functions we fit, hence getting a value of $\R_0$ that converges to it does not take long.
  \item Once we converge to the annual attractor, we open the \texttt{AUTO} window in \texttt{XPPAUT} and follow the annual attractor across values of $\R_0$.
    In most cases, the annual attractor will become a repeller and branch off into a biennial attractor.
    We fix $\R_0$ to this value at the period doubling bifurcation.
    It is possible that the period doubling bifurcation does not exist, in which case the annual attractor continues for all $\R_0$ and hence we cannot find a respective $\alpha$ value for the sinusoidally forced model.
    In this scenario, we assume that the model is predicting only period $1$ attractors, which is plausible given that in \cref{fig:TwoPar}, for a given $\alpha$, if the period $2$ attractor does not exist, then neither do any non-period $1$ attractors.
  \item Now, we set up a two parameter bifurcation diagram on $\alpha$ and $p$.
    We follow the value of $\alpha$ that keeps the period doubling bifurcation fixed across $p$, until $p=1$.
    At $p=1$, we obtain a value of $\alpha$ such that the corresponding sinusoidally forced model has the same period-doubling bifurcation in $\R_0$ as the \texttt{macpan2}-forced model, which~\citet{PapsEarn19} imply will yield nearly identical dynamics.
    We display an example two-parameter bifurcation diagram in \cref{fig:alpha_cont}.
  \item We repeat this procedure for all other shape functions from \texttt{macpan2}.
\end{enumerate}

\begin{figure}[p]
  \centering
  \includegraphics[height=0.58\textheight,keepaspectratio]{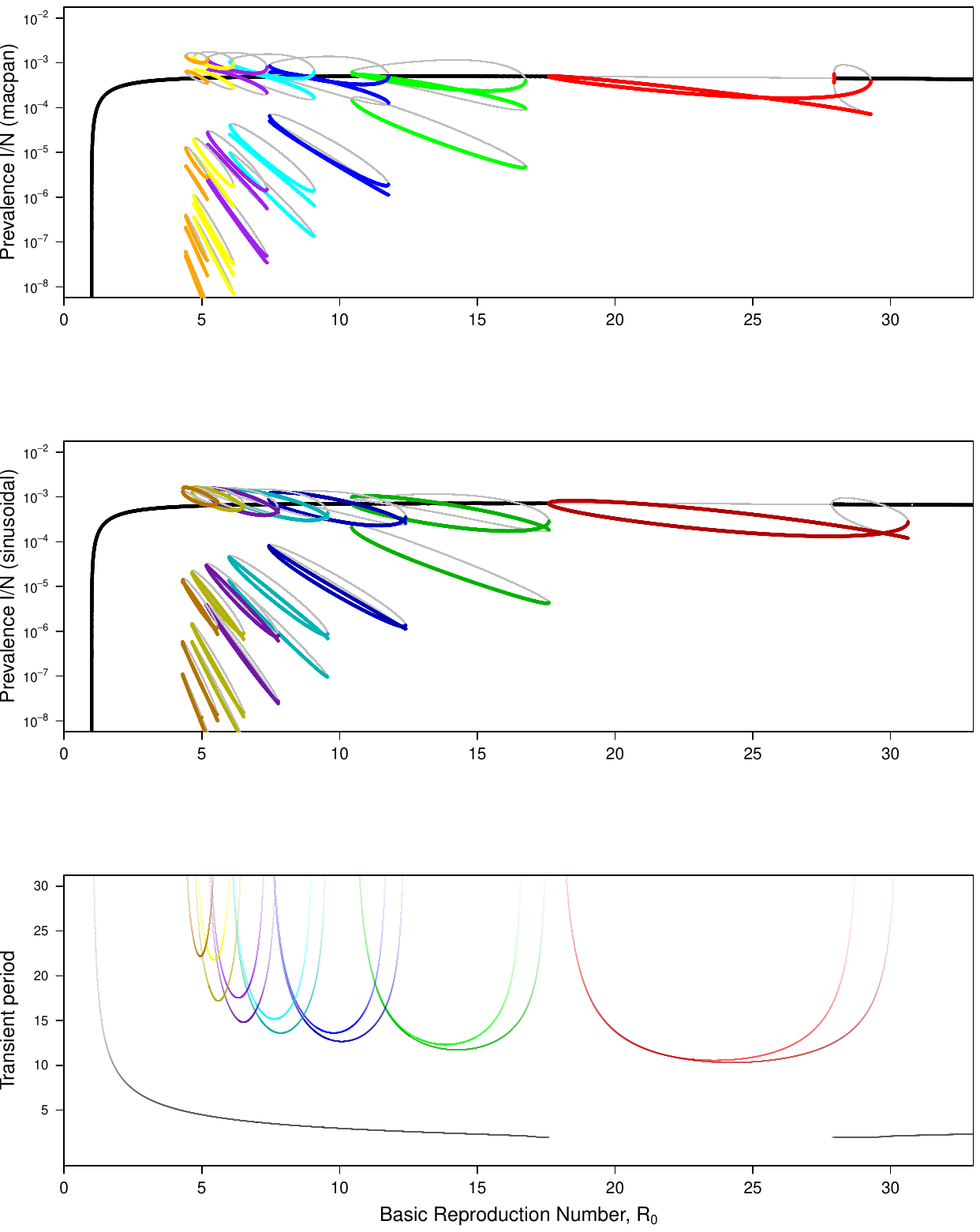} \caption{Top: Asymptotic dynamics of the SIR model for scarlet fever ($\gamma^{-1} = \SI{15}{days}$, and $\nu = \mu = \SI{0.02}{yr^{-1}}$) using average fitted \texttt{macpan2} forcing from between 1870 and 1880.
    Black, red, green, blue, cyan, magenta, and yellow lines represent attractors of integer periods from 1 to 7, respectively.
Thin lines represent the respective repellers.
Middle: Asymptotic dynamics of the SIR model using sinusoidal forcing, with appropriate $\alpha$ being determined from the previous \texttt{macpan2} forcing via the methods of~\cite{PapsEarn19}.
From their procedure, we obtained $\alpha \approx 0.11$, which indeed yields similar dynamics across $\R_0$.
Bottom: Transient periods of the respective attractors for both the \texttt{macpan2} (lighter) and sinusoidally (darker) forced models.
These results were computed using \texttt{XPPAUT} \citep{Erme02}, following the methods described in the Supplementary Materials of~\cite{KrylEarn13} to numerically calculate the periodic attractors.}\label{fig:AttTran}
\end{figure}

\begin{figure}[p]
	\centering
	\includegraphics[height=0.56\textheight,keepaspectratio]{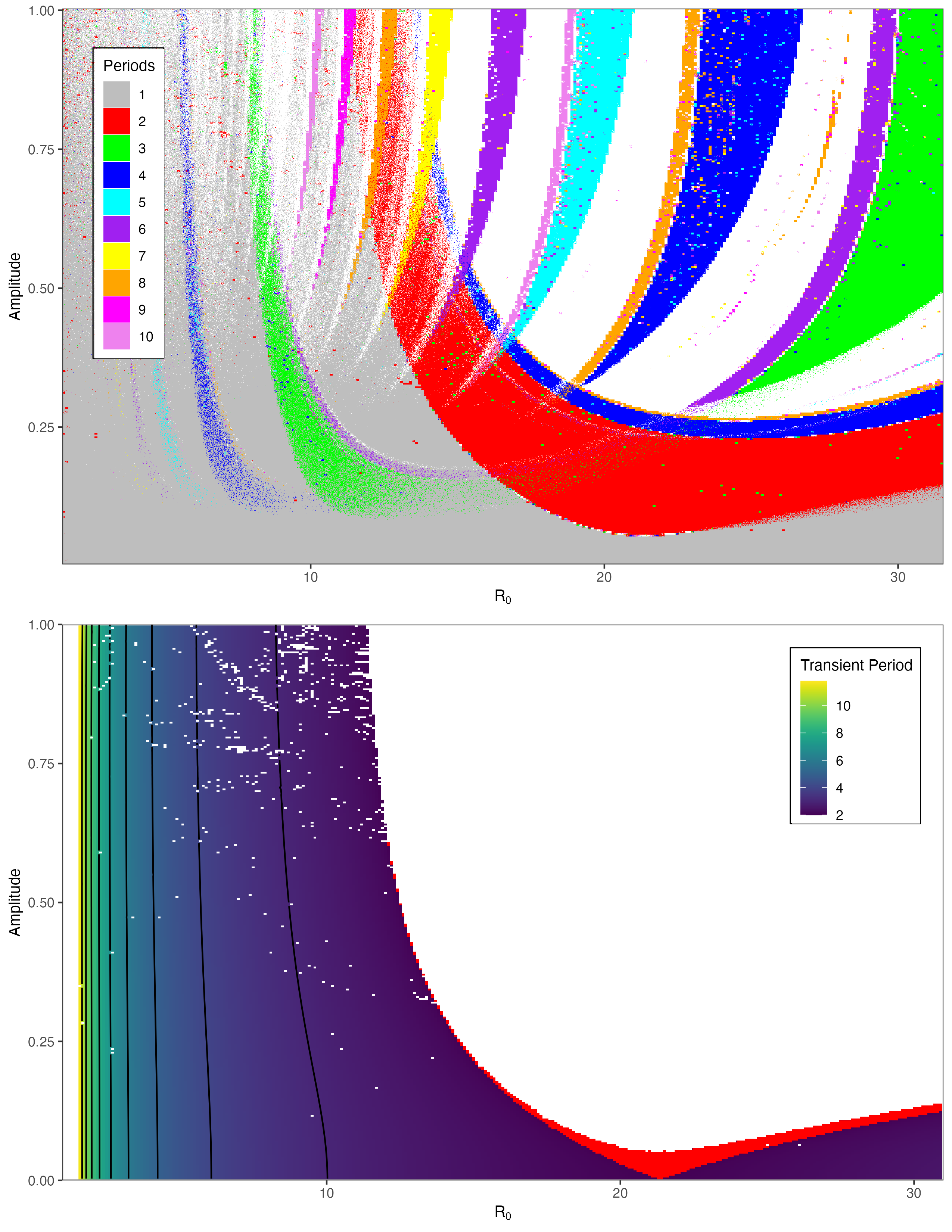} \caption{Top: Attractors of the sinusoidal SIR model identified using the procedure outlined above.
Since each point of the $300\times 300$ grid contains 100 periods, we split the squares of the grid into $10\times10$ sub-grids, randomly mapping each of the 100 periods to a point in the sub-grid, allowing the relative size of the basins of attraction to be observed from this plot.
Note that periods of $2k$ appearing on the tail end of period $k$ curves could be due to the model needing longer than 1000 years to converge near bifurcations.
Bottom: Transient periods associated with the annual attractors of the sinusoidal SIR model (across $2,521,899$ ``valid'' parameterizations).
Contour lines are displayed at integer periods.
The area in red represents transient periods that were estimated to equal $2$, this phenomenon has been observed in the seasonally forced SIR model when given the parameters of various diseases \citep{BaucEarn03b}.}\label{fig:TwoPar}
\end{figure}

No period-doubling bifurcation was found for 1892--1899 or 1924--1939.  We therefore could not obtain an equivalent $\alpha$ for these intervals, in which the model predicted only annual attractors.
With an equivalent $\alpha$ identified for each of our shape functions, we can now analyse the dynamics over time across $(\R_0, \alpha)$ space.
We can quantify these dynamics by looking at the model's attractor and transient periods.
The attractor periods are obtained from the orbits that the model converges to asymptotically across various initial conditions.
These attractor periods will always be an integer multiple of 1 year due to the period of forcing \citep{Earn+00}.
The transient periods are not restricted to only integer multiples, and represent the period of dampened oscillations as the model converges to an attractor.
The transient period $T_k$ for any periodic attractor of period $k$ is $2\pi k/|\arg(\lambda_k)|$, with $\lambda_k$ being the dominant eigenvalue of the stroboscopic map \citep{HempEarn15,BaucEarn03a}.
Note that we expect to observe transient periods in both mortality and case data, as demographic stochasticity such as random births and deaths constantly push the system away from the attractor \citep{BaucEarn03a}.
We display the attractor and transient periods for the model with \texttt{macpan2} forcing and equivalent sinusoidal forcing from the part of the transmission rate between 1870 and 1880 in \cref{fig:AttTran} as an example.
From \cref{fig:AttTran}, we can see that the procedure outlined above is effective at maintaining both transient and attractor periods, with deviations only in the transient periods of non-annual attractors.
Only the transient period of the annual attractor is short enough to be observed in our time series, hence the deviations in other transients are not a concern.

Although we could repeat these methods for each part of our partitioned time series, we would like to be able to observe the dynamics of the model across both $\alpha$ and $\R_0$, not just across $\R_0$ for various fixed $\alpha$ values.  
Our methods provide a finite set of $\alpha$ values that jumps between values at the break points of our partitions, however realistically, we would expect $\alpha$ to be a smooth function over time, hence knowing the behaviour of the model in between these jumps can be useful.
We use the following methods inspired by~\citet{HempEarn15} to simulate all the possible attractor and transient periods across $(\R_0, \alpha)$-space and estimate the size of the basins of attraction for the values of $\R_0$ and $\alpha$ that our model predicts over time:
\begin{enumerate}
	\item We created a $300\times300$ grid in $(\R_0, \alpha)$-space by taking 300 evenly spaced values for both $\R_0 \in [1.6, 31.5]$ and $\alpha \in [1/300, 1]$.
	\item At each point in the grid, we simulated the SIR model for $1000$ years, with a time-step of $0.001$ years (to give the model enough time to converge).
    As before, we took $\nu = \mu = \SI{0.02}{yr^{-1}}$ and $\gamma^{-1} = \SI{15}{days}$.
    This was done across $10\times10$ evenly spaced values of $S_0$ and $I_0$, with $S_0 \in [0.8/\R_0, 1.2/\R_0]$ (as $1/\R_0$ is the proportion of susceptibles in the unforced model at equilibrium), and $I_0 \in [0.0002/10, 0.0002\cdot10]$ (See \cref{appendix:InitialValues} for an explanation of why we choose prevalence/$I_0$ near $0.0002$).
This gives us an estimate for the sizes of the basins of attraction.
	\item With each simulated solution, we identified how many unique values of $I$ were present at $t = 950, 951, 952, \dots, 1000$ years.
This tells us the length of the cycle the model converged to, hence the attractor period.
For cycles of length $>10$, we assumed that the model failed to converge, and hence assigned them a missing value.
\end{enumerate}
We completed this procedure, and plot the results in the top panel of \cref{fig:TwoPar}.

We also estimated the transient periods associated with the annual attractors, which were the only transients observable for the length of time series we are analyzing.
To estimate this transient period, we estimated the Jacobian of the stroboscopic map once converged at the $1$ year cycle by taking $10$ equally spaced points in a small radius ($r = 10^{-5}$) around the attractor in $(S,I)$-space.
We can then estimate partial derivatives of the stroboscopic map by simulating the model starting at these points for $1$ year and at points offset by a small distance ($10^{-6}$) in the $S$ or $I$ direction.
From these partial derivatives, we construct $10$ Jacobians, each of which we can use to obtain an estimate of $\lambda$, the dominant eigenvalue, and hence we also obtain $10$ estimates of $T_1$.
We estimate $T_1$ at each set of parameters and initial conditions where the annual cycle exists (``valid'' parameterizations) by taking the average of the $10$ estimates each of these parameterizations produce.
These $10$ estimates are generally consistent with one another: across all annual-attractor transients computed this way, the upper 95\% quantile of their coefficient of variation is $0.037$, and $T_1$ is always at least $2$ years.
We plot the resulting transient periods in the bottom panel of \cref{fig:TwoPar}.

With these two-parameter attractor and transient diagrams, we can plot the trajectory our model took in $(\R_0, \alpha)$-space over time, using the respective time-varying parameters.
We observe from \cref{fig:TwoParTraj} that for intervals in which we could obtain $\alpha$, only annual attractors are predicted for 1842--1857, 1880--1892, and 1899--1924, with some potential for period-5, period-6, and period-7 attractors between 1857 and 1880.
For these same time periods, we predict transient periods between $3.4$ and $6.4$, primarily driven by changes in $\mathcal{R}_0$.
We plot the basins of attraction for the various attractors that the sinusoidal model predicts for the time series, and compare them to the periods observed in the wavelet spectrum of the observed mortality and incidence series \cref{fig:Basin}.
\section{Results}

From the sinusoidally forced SIR model, we correctly predict the annual attractor before 1892 in the mortality series, and from 1901 to 1924 in the incidence series.
Our model predicts the annual cycle to always exist, though we can see from the wavelet spectra that this does not seem to be the case in the mortality series.
The disappearance of the annual cycle in the mortality series is likely due to the low number of deaths, which would have masked the signal from the annual cycle.
Based on the wavelet spectrum of the incidence series, it seems highly plausible that even before 1901, there were still consistent yearly epidemics of scarlet fever.
Attractors of periods 5, 6, 7, and 8 are found to be highly unlikely (in less than 3\% of nearby parameterizations) between 1857 and 1880.
These correspond well with the longer periodicities observed in the mortality series at this time.  Although convergence to these attractors is unlikely, proximity to a periodic orbit in parameter space can produce transient dynamics that resemble the nearby cycle even after the attractor no longer exists \citep{RandWils91,Earn+00}.

We find that many of the other observed periods can be explained by the transient period:
From 1842 to roughly 1865, the transient period can explain the approximately 3.5-year periodicity in the mortality series.  Although ambiguous in the wavelet spectrum, a four-year cycle may be present from 1865 to 1870.  After 1870, the predicted transient periods closely match the observed long periodicities as both lengthen over time.
Although we cannot estimate the transient period precisely when no equivalent $\alpha$ is available (1892--1899 and 1924--1939), \cref{fig:TwoParTraj} indicates that it is driven mainly by $\R_0$, with increasing $\R_0$ producing shorter transients.  Because the fitted $\R_0$ changes slowly, we infer that the transient period remains close to 4.5 years during 1892--1899 and continues to lengthen after 1924 as $\R_0$ declines.
Both of these inferences would accurately predict longer periods observed in the mortality series.

Our model does not explain the five-year pattern before 1857 or the approximately three-year pattern between 1887 and 1897.
There may also be an unexplained two-year cycle from 1842 to roughly 1858 and an approximately four-year cycle after 1925, although both are ambiguous in the wavelet spectrum.
\section{Discussion and Conclusion}
Based on our results, the asymptotic and transient dynamics of the SIR model are able to explain nearly all the transitions observed in the scarlet fever mortality series.
We were successful in explaining many of the dynamics in the scarlet fever mortality and incidence series, such as longer periods of approximately $\SI{4}{yr}$ being the result of transients and attractors.
One concern in our analysis is that, as discussed by~\citet{Hard93}, the expected lethality and infectiousness of scarlet fever depend on the dominant strain, which we infer to have changed during our time series from the observed changes in mortality rates.
Although we accounted for the lower mortality rate of the disease as the dominant strain changed once around 1880, we cannot be certain that this was the only time there was a shift in the dominant strain.
Having more changes in mortality rates can impact our inferences, as our fitted transmission rate trend is dependent on $\CFP$, hence changes in $\CFP$ could lead to different $\R_0(t)$ curves, in turn predicting vastly different dynamics.

Scarlet fever can also be acquired more than once by the same host,
contrary to the SIR assumption that infection confers lifelong
immunity.  A model with loss of immunity could return some recovered
individuals to the susceptible class, but the required historical
reinfection rates are unavailable.  A modern clinical report found
recurrence in 10--15\% of patients while noting that recurrence had
historically been considered uncommon \citep{deDi17}.
Furthermore, since it is known that the dominant strain of scarlet fever has changed in the past, modern studies on scarlet fever reinfection are likely not applicable to the strains from our time series in London, hence making it near impossible to estimate the effects of reinfection with just the data available.

The model also introduces births directly into the susceptible class,
thereby ignoring temporary protection from maternal antibodies during
infancy.  This delay is short relative to the multi-year periodicities
that are the focus of our transition analysis, but it is another way
in which the model simplifies the immunological natural history of
scarlet fever.

Comparing our results to~\cite{HempEarn15}, who conducted a similar analysis on measles in New York City, we find several similar patterns between our \cref{fig:TwoPar} and their Figure~7.
This is to be expected, as the parameters used are very similar between measles and scarlet fever (mean generation interval of 13 and 15 days, with birth and death rates set to $\SI{0.02}{yr^{-1}}$).
Looking at Figure~9 of~\cite{HempEarn15}, their model is similar to ours in that both accurately predict the annual cycle for most of the time series, which is to be expected since both measles and scarlet fever are childhood diseases impacted by seasonal factors \citep{HeEarn16}.
Both analyses are also able to explain longer periods (2--3 years in~\cite{HempEarn15}, $\sim4$ years for us) as the result of transients and attractors (biennial cycle in \cite{HempEarn15}, 5--6 year cycles for us).

\begin{figure}
	\centering
	\includegraphics[height=0.56\textheight,keepaspectratio]{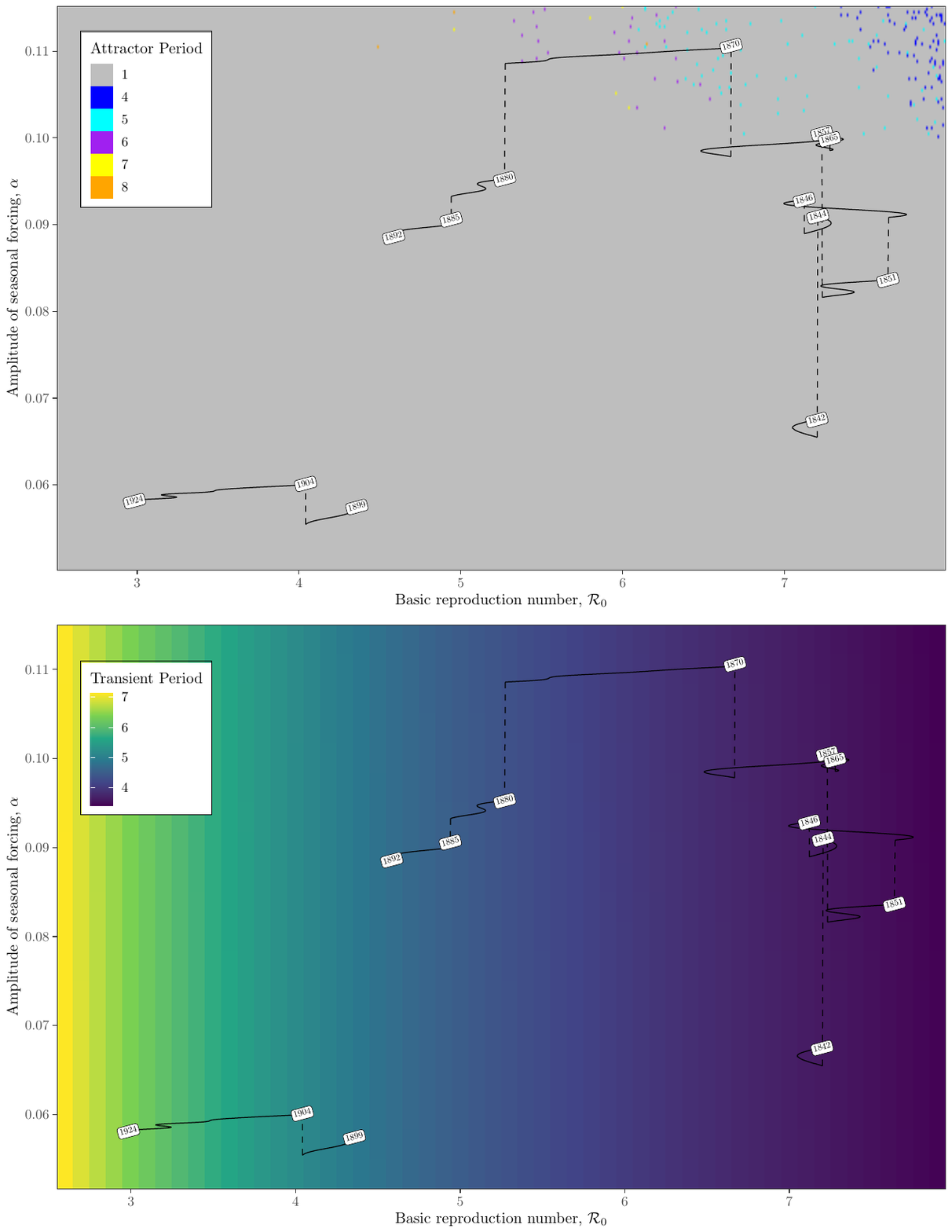} \caption{Magnified version of \cref{fig:TwoPar} with labelled trajectories of the estimates for $\alpha$ and $\R_0$ for the scarlet fever mortality time series.
Note that white squares represent simulations that did not converge to a period $< 10$.
Dotted vertical jumps represent instantaneous changes in amplitude at break point years.
Text boxes represent the beginning of each break in amplitude, with the solid line showing the trajectory.
Amplitude remains constant between break years, hence the plot of the actual estimates would have all solid lines be horizontal.
To make the change in $\R_0$ between break years clear (as constant $\alpha$ would produce  horizontal lines from which the trajectory of $\R_0$ cannot be determined), we plot the estimate of $\alpha$ at each break year, and then have it decrease until the next break year, when in reality, $\alpha$ is constant between jumps.}\label{fig:TwoParTraj}
\end{figure}

\begin{figure}[p]
	\centering
	\includegraphics[height=0.58\textheight,keepaspectratio]{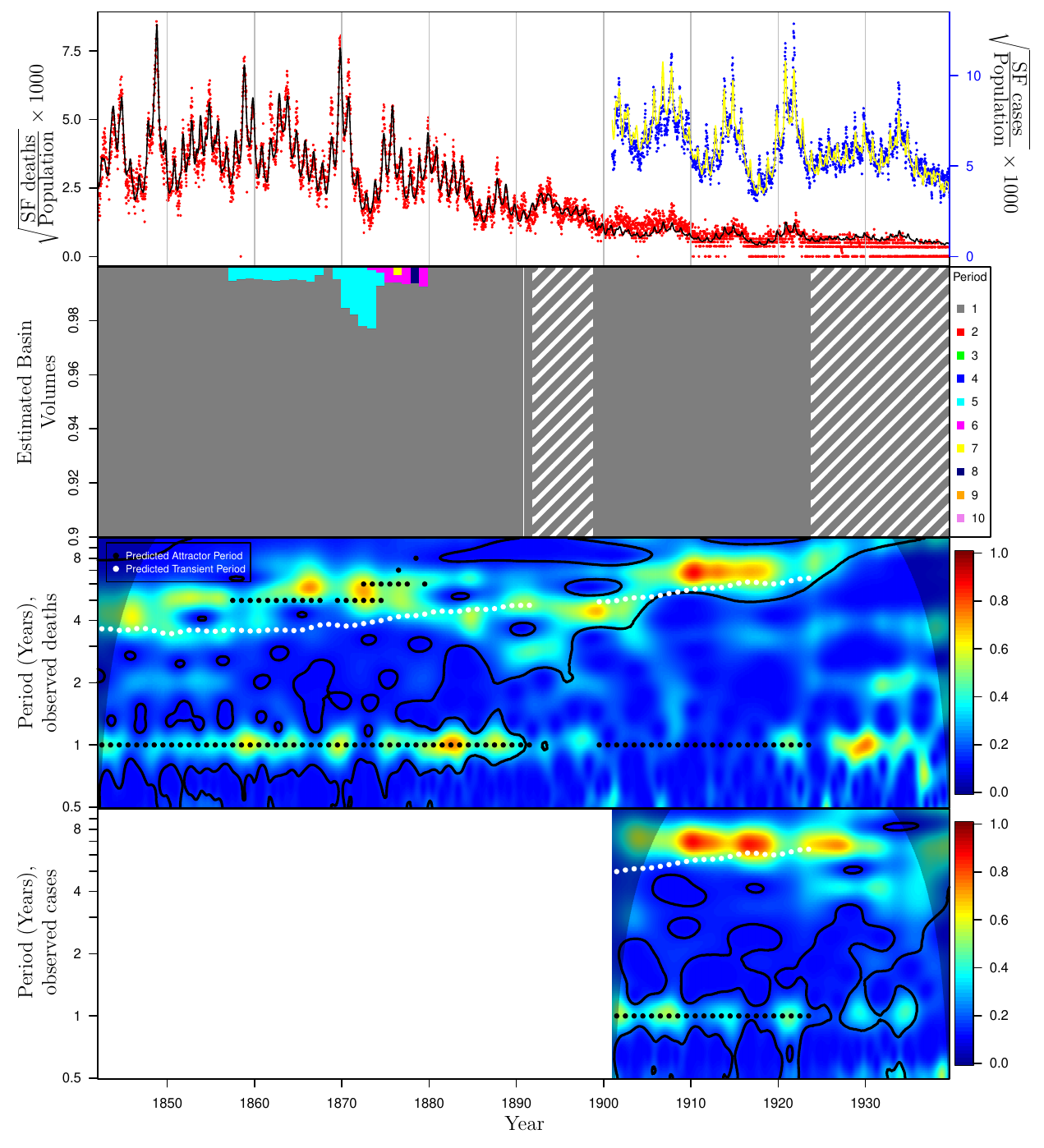} \caption{\footnotesize Top: Square root of scarlet fever mortality series normalized by all-cause mortality.
Top middle: Relative sizes of the basins of attraction of the model given the time-varying parameters inferred from the time series.
Basin sizes were estimated using 10000 simulations each year with $S_0$ and $I_0$ being varied.
White space represents initial conditions where the model failed to converge to a period of $<10$.
Prior to 1898, more than 90\% of simulations converged to the annual attractor, hence the y-axis only displays between $.9$ and $1$.
Bottom middle and bottom: Wavelet spectrum of the square root ACM-normalized observed/fitted scarlet fever mortality series respectively, with 95\% confidence contours shown in black.
Obtained using the \texttt{WaveletComp} R package \citep{RoesSchm18}.
Attractors and transients associated with the annual attractor based on the sinusoidally forced SIR model are displayed with points where applicable.}\label{fig:Basin}
\end{figure}

We can explain observed transitions in the wavelet spectra as a result of mechanistic changes in the underlying parameters of our SIR model.
For example, \cref{fig:Basin} shows that the longer periodicity, which increases from about 3.5 years in 1842 to about 7--8 years around 1920, corresponds to the transient period of the annual attractor.  We therefore infer that its evolution was driven mainly by changes in $\R_0$, because \cref{fig:TwoParTraj} shows that this transient period depends almost entirely on $\R_0$ in the relevant region of parameter space.
Specifically, the transient period grows longer over time since $\R_0$ was decreasing over time, which mainly appears to be driven by a steady decrease in $\beta_0$ over time, meaning a steady fall in the transmissibility of scarlet fever.
From asymptotic analysis, aside from the predicted annual cycles, our model could accurately predict longer integer periods between 5 and 8 years between 1857 and 1880, which is consistent with the observed longer periods in the mortality series.
From the top panel of \cref{fig:TwoParTraj}, this can be attributed to a greater amplitude in seasonal variance alongside a greater $\R_0$ value ($\R_0 > 5$) during this time.
The larger value of $\R_0$ can be attributed to higher birth rates at this time (as seen in \cref{fig:BirthSeries}), whereas larger seasonal variance is more difficult to explain mechanistically, though could be due to changes in the strain of scarlet fever near this time.
These explanations differ from those of \citet{Earn+00}, who found that changes in birth rates were sufficient to explain the transitions in recurrent measles epidemics in both the United States and England and Wales.

In previous work, such as \citet{Earn+00} and \citet{HempEarn15}, inferences could only be made based on observed data, such as changes in birth and vaccination rates, however with our \texttt{macpan2} methodology, we now have a way of estimating unobserved time series, such as case-fatality or changes in transmission rate patterns, from a self-consistent fit of the model.
Whereas these previous studies considered incidence data, our analysis extends transition analysis to the joint fitting of incidence and mortality data.
These methods can be applied to many other infectious disease data sets to provide insights into changes in their infectiousness patterns.

  \section*{Declaration of AI use}
  \msaiuse

  \section*{Funding}
  \msfunding

  \section*{Acknowledgements}
  \msacknowledgements

\newpage
  \bibliographystyle{chicago}
  \bibliography{Thesis_Refs}
\newpage
\appendix
\section{Appendix: Starting values for parameter estimates}\label{appendix:InitialValues}

\cref{tab:ParamTable} lists the parameters we estimate, including the starting values
we chose, and the final maximum likelihood estimates we obtained.

The initial susceptible proportion, $S_{0}$, represents the proportion of the total population assumed to be susceptible at the first time step, which we chose to be $1/6.5$, as~\citet{AndeMay91} estimated $\R_0 \approx 6.5$, and $1/\R_0$ is the value $S$ converges to in the unforced SIR model.
We assume that at the beginning of the time series, the system was near equilibrium.
For the infected proportion/prevalence, $I_{0}$, we based our initial value on case rates reported by \citet[Table~3.5, p.\,66]{Hard93} from London County Council archival records.
These yearly case rates between 1891 and 1900 inclusive ranged from $2.7$ to $8.6$ per 1,000 population, with mean $4.83$ per 1,000.
Assuming the case-rate at the beginning of the time series was near this value, we multiplied this mean value by the mean generation interval of $15/365.25$ years to get an estimate of the prevalence of $0.000198 \approx 0.0002$.
\cite[Table~3.5, p.\,66]{Hard93} also included case fatality rates of notified cases between 1891 and 1900, ranging from 2.2\% to 5.1\%.
This was after the milder strain of scarlet fever became dominant.
We assume that these values overestimate the CFP throughout the time series, as these were based on notified cases, hence many milder cases were likely underreported.
Hence $\text{CFP}_{\rm min}$, the minimum value of our CFP logistic function, was initialized to $0.01$ and $\text{CFP}_{\rm max}$ was initialized to $0.025$ (meaning the CFP of scarlet fever starts around $0.025$, and then approaches $0.01$ later in the time series).
For $\phi_M$ and $\phi_C$, the dispersion parameters of the negative binomial distribution we used to model observations of scarlet fever deaths and cases, we used an initial value of $e$ since we expected the data to be overdispersed (hence $\phi > 1$), and since we were using a log link function, we chose a simple value of $1$ under the link function.
The rate parameter for the CFP logistic function, $\CFP_{\rm rate}$, was initialized to $0.01$ as an arbitrary value.
The time of inflection for the CFP curve, $t_{\rm inf}$, was initialized to be the time step in the middle of our model.
This includes the 479 weeks of the mortality series we repeat at the beginning and 508 weeks at the end.
This value should be appropriate, since it is close to 1890, when the disease was falling significantly in mortality.
The baseline parameter for the trend of $\beta$, $b_0$, where $\beta_{\rm trend}(t) = \exp(b_0 + \mathbf{F}(t)\cdot\mathbf{c})$, was initialized to $0$, as we had no compelling evidence for what this value should be to produce accurate $\beta(t)$ estimates.
The standard deviation parameters for the weights, $\sigma_\beta$ and $\sigma_{\rm f}$, were initialized to $1$ since it seemed like a natural choice to model these weights under a standard normal distribution.
Lastly, the weight parameters, $c_i, a_{i,j}, b_{i,j}$, were initialized to $0$ as they were assumed to have mean $0$, and could be either negative or positive.

For the link functions, all the proportions, $S_{0}, I_{0}, \text{CFP}_{\rm min}, \text{CFP}_{\rm max}$, used a logit link so they would be between $0$ and $1$.
$\phi$, $\CFP_{\rm rate}$, $t_{\rm inf}$, $\sigma_\beta$, and $\sigma_{\rm f}$ used a log link function since these parameters were all strictly positive.
Parameters that could plausibly be negative and/or positive, $b_0$, $c_i$, $a_{i,j}$, and $b_{i,j}$, used an identity link function.

We performed a sensitivity analysis on the initial values of the CFP parameters, $\text{CFP}_{\rm min}$, $\text{CFP}_{\rm max}$, $\text{CFP}_{\rm rate}$, and $t_{\rm inf}$, as these were the values we had the least confidence in.
Each parameter was doubled, kept identical, or halved (except for $t_{\rm inf}$, which we varied by $\pm 10\%$ or kept identical), meaning each of the four CFP parameters had three possible initial values, giving us $3^4 = 81$ combinations of initial values to test.
All $81$ combinations converged to the same parameterization as the original initial values, supporting the robustness of the estimates to the starting values of the CFP parameters.

\FloatBarrier
\newpage

\centerline{\huge\bfseries TABLES}

\bigskip \bigskip

\begin{table}[p]
\caption{Estimated incubation, latent, and infectious periods of scarlet fever, given in days.
  The data from~\cite{AndeMay82} are from epidemics in England and Wales between 1897 and 1978; the remaining estimates are from textbooks that do not identify their primary sources.}
\centering
  \begin{tabular}{cccc}
\toprule
Incubation period & Latent period & Infectious period & Source \\
\midrule
2--3               & 1--2          & 14--21 & \citep[p.\,1055]{AndeMay82}\\
1--3               & ---             & 10--21 & \citep[p.\,553]{Heym04} \\
1--4               & 1--2          & ---    & \citep[p.\,120]{Rose18} \\
1--5               &  ---            & 7--21 & \citep[p.\,833]{Shar+16} \\
\bottomrule
\end{tabular}\label{tab:Params}
\end{table}

\begin{table}[p]

\caption{\label{tab:ParamTable}Parameters fitted jointly to the scarlet fever mortality
           and case series by maximum likelihood, with starting values chosen
           as described in Appendix \ref{appendix:InitialValues}.  The
           case-fatality proportion was represented by a time-varying logistic
           curve, and deaths and cases were modelled with separate negative
           binomial observation distributions.  Radial-basis and Fourier
           weights were penalized using zero-mean normal densities with
           standard deviations $\sigma_{\beta}$ and $\sigma_f$,
           respectively.  We report only the ranges of the fitted weights.}
\centering
\footnotesize
\begin{tabular}[t]{l>{\raggedright\arraybackslash}p{1.5in}ll>{\raggedright\arraybackslash}p{0.8in}>{\raggedright\arraybackslash}p{0.6in}}
\toprule
Symbol & Parameter description & Initial value & Fitted value & Link function & Effect type\\
\midrule
$S_{0}$ & Initial susceptible proportion & $0.154$ & $0.123$ & Logit & Fixed\\
$I_{0}$ & Initial infected proportion & $0.0002$ & $0.000594$ & Logit & Fixed\\
$\text{CFP}_{\rm min}$ & Minimum case fatality proportion & $0.01$ & $0.0141$ & Logit & Fixed\\
$\text{CFP}_{\rm max}$ & Maximum case fatality proportion & $0.025$ & $0.0403$ & Logit & Fixed\\
$\phi_M$ & Dispersion parameter of negative binomial for mortality & $2.72$ & $17.6$ & Log & Fixed\\
\addlinespace
$\phi_C$ & Dispersion parameter of negative binomial for cases & $2.72$ & $31.14$ & Log & Fixed\\
$t_{\rm infl}$ & CFP logistic curve time of inflection & $1890.86$ & $1882.17$ & Log & Fixed\\
$\text{CFP}_{\rm rate}$ & CFP logistic curve rate parameter & $0.01$ & $0.05$ & Log & Fixed\\
$b_0$ & Baseline value for mean $\beta$ & $0$ & $0.805$ & Identity & Fixed\\
$\sigma_{\beta}$ & Standard deviation of $\beta$ weight parameters & $1$ & $0.32$ & Log & Fixed\\
\addlinespace
$\sigma_{\rm f}$ & Standard deviation of Fourier weight parameters & $1$ & $0.02$ & Log & Fixed\\
$c_i$ & $\beta$ weight parameters & $0$ & $[-0.82,0.67]$ & Identity & Fixed\\
$a_{i,j}$ & Fourier cosine weight parameters & $0$ & $[-0.071,0.028]$ & Identity & Fixed\\
$b_{i,j}$ & Fourier sine weight parameters & $0$ & $[-0.05,0.085]$ & Identity & Fixed\\
\bottomrule
\end{tabular}
\end{table}
\begin{table}[p]
	\caption{Known parameters used for scarlet fever modelling in \texttt{macpan2} between 1842 and 1939 in London, England.} \label{table:pars}
  \centering
	\begin{tabular}{p{0.1\linewidth}p{0.2\linewidth}p{0.325\linewidth}p{0.13\linewidth}p{0.1\linewidth}}
    \toprule
		Parameter &
		Parameter type &
		Meaning &
		Estimate/

    Range &
		Source \\ 
    \midrule
		$\gamma^{-1}$ &
		Fixed &
		Mean generation interval &
		15 days &
		\cite{Rose18}, \cite{Shar+16}, \cite{Heym04}, \cite{AndeMay91} \\
    \addlinespace
		$N$ &
		Time-varying &
		Weekly population of London, found by linear interpolation between decennial population estimates &
		2,253,055--8,614,252 &
		\cite{Demographia} \\
    \addlinespace
		$B$ &
		Time-varying &
		Trend in weekly births, found by using empirical mode decomposition on weekly births &
		1090--2596 &
		RGWR \\
    \bottomrule
	\end{tabular}
\end{table}

\FloatBarrier
\newpage

\centerline{\huge\bfseries FIGURES}

\bigskip \bigskip

\end{document}